\documentclass[aps,
 reprint,
nofootinbib,
 amsmath,amssymb,
 aps,superscriptaddress,
]{revtex4-1}

\usepackage{graphicx}
\usepackage{dcolumn}
\usepackage{bm}
\usepackage{lipsum}
\usepackage{amsmath,amssymb}
\usepackage{caption}
\usepackage{subcaption}
\usepackage{booktabs}
\usepackage[utf8]{inputenc}
\usepackage[T1]{fontenc}
\usepackage[normalem]{ulem} 
\usepackage{verbatim}
\usepackage[left=2cm, right=2cm, top=2cm]{geometry}

\usepackage{epigraph}
\usepackage{siunitx}
\usepackage[english]{babel}
\usepackage{listings}
\usepackage{graphicx}
\usepackage[utf8]{inputenc}
\usepackage{listings}
\usepackage{matlab-prettifier}
\usepackage{xcolor}
\usepackage{amsfonts}
\usepackage{amsmath}
\usepackage{amssymb}
\usepackage{amsthm}
\usepackage{array}
\newcolumntype{P}[1]{>{\centering\arraybackslash}p{#1}}
\newcolumntype{M}[1]{>{\centering\arraybackslash}m{#1}}
\usepackage{subcaption} 
\usepackage{ragged2e}
\DeclareCaptionJustification{justified}{\justifying}
\usepackage{subcaption}
\usepackage{epigraph}
\usepackage{csquotes}
\usepackage{braket}
\usepackage{xcolor}
\usepackage{enumitem}   
\usepackage[colorlinks = true,
            linkcolor = blue,
            urlcolor  = blue,
            citecolor = blue,
            anchorcolor = blue]{hyperref}
\usepackage{comment}
\usepackage{hyperref}
\usepackage{color}

\newcommand{\KIAA}{\affiliation{Kavli Institute for Astronomy and Astrophysics, Peking University, Beijing 100871, China}}
\newcommand{\DOA}{\affiliation{Department of Astronomy, School of Physics, Peking University, Beijing 100871, China}}
\newcommand{\LdIT}{\affiliation{Laboratoire des 2 Infinis - Toulouse (L2IT-IN2P3), Université de Toulouse, CNRS, UPS, F-31062 Toulouse Cedex 9, France}}
\newcommand{\Ins}{\affiliation{Dipartimento di Scienza e Alta Tecnologia, Università dell’Insubria, via Valleggio 11, I-22100 Como, Italy}}

\begin{document}
\title{Strongly-lensed extreme mass-ratio inspirals}
\author{Martina Toscani}
\email{martina.toscani@l2it.in2p3.fr}\LdIT
\author{Ollie Burke}
\email{ollie.burke@l2it.in2p3.fr}\LdIT
\author{Chang Liu}
\email{leslielc@pku.edu.cn}\DOA\KIAA\LdIT
\author{Nour Bou Zamel}\LdIT
\author{Nicola Tamanini}\LdIT
\author{Federico Pozzoli}\Ins

\date{\today}

\begin{abstract}
In this work, we investigate detection rates and parameter estimation of strongly-lensed extreme mass-ratio inspirals (LEMRIs) in the context of the Laser Interferometer Space Antenna (LISA).
Our results indicate that LEMRIs constitute a new gravitational-wave target signal for LISA, with detection rates ranging from zero to $\sim 40$ events over a four year-observation period, and that it is possible to reveal and characterize LEMRIs at redshift $z \gtrsim 1$. We finally show that one LEMRI observation with identified host galaxy may yield percent constraints or better on the Hubble constant.
\end{abstract}

\maketitle

\section{\label{sec:level1}Introduction}
The most massive galaxies in our Universe are expected to host a massive black hole (MBH, $10^{4} - 10^{9} \text{M}_{\odot}$) in their core \citep{Genzel:10aa,Lyndel:71aa,Soltan:82aa},  surrounded by nuclear star clusters of millions of stars \citep{Scholder:14aa}. While these stars could be disrupted due to tides induced by the MBH gravitational field and thus produce both electromagnetic (EM, see recent reviews \citep{Vanvelzen:20aa, Saxton:20aa} and references therein) and gravitational wave (GW) emission~\citep{Toscani:22aa, Toscani:23aa}, compact objects (COs) such as stellar-mass black holes and neutron stars survive the tidal forces of the MBH and start inspiralling towards it.
The inspiral of a $\sim 1-100 \text{M}_{\odot}$ CO into a $10^{4} - 10^{7}\text{M}_{\odot}$ MBH is called an extreme mass-ratio inspiral (EMRI).
The formation mechanisms of EMRIs are still largely uncertain but recent studies find that the Laser Interferometer Space Antenna (LISA)~\citep{Lisaconcept:17aa,Seoane:23aa} may be able to detect $\sim 1 - 1000$ EMRIs per year, depending on the underlying astrophysical assumptions of different EMRI population models \citep{gair2004event, gair2017prospects, Babak:17aa, Pozzoli:23aa}.
Computing EMRI waveforms is challenging, requiring tools from black hole perturbation theory to accurately describe the behaviour of the CO within the strong-field regime of the central MBH~\citep{teukolsky1972rotating, bardeen1972rotating, detweiler2001radiation, barack2009gravitational, pound2012second, barack2018self, gralla2008rigorous, drasco2006gravitational, babak2007kludge}. The specific characteristics of the CO orbital dynamics are sensitively encoded in the GW, and, due to the large number ($\sim 10^{4} - 10^{6}$) of observable orbital cycles, EMRI observations offer unparalleled precision constraints on parameters that govern the system~\citep{Babak:17aa, Burke:20aa, piovano2021assessing, huerta2009influence, speri2021assessing}. Finally, EMRIs provide an excellent laboratory to perform tests of strong-field gravity \citep{gair2013testing, chua2018towards, yagi2016black, maggio2021extreme, canizares2012testing, hughes2006sort, barack2007using, glampedakis2006mapping}.\\
\indent To accurately retrieve the parameters of an EMRI, and consequently correctly characterize the observed EMRI population, it is crucial to understand how gravitational lensing can impact the detected signals. 
Gravitational lensing (see, e.g.,~\citep{Ohanian:74aa, Schneider:92aa}) occurs when a massive object -- \textit{the lens} -- along the line of sight between the observer and the source, bends the surrounding space-time, deflecting the signal from its original path. In particular, we refer to strong lensing when we have the production of magnified multiple images of the same source \citep{Ng:18aa, Oguri:18aa, Pang:20aa,Haris:18aa,Hannuksela:19aa,Li:23aa,Broadhurst:19aa,McIsaac:20aa,Liu:21aa}, which arrive at the detector at different times. If the duration of the signal is longer than the typical time delay between two images, as in the case of EMRIs, these will superimpose in time in the LISA data stream. The ability to recognize lensed events in the data is crucial to unveil correct information regarding the GW source and the lens, and also to prevent us from a biased reconstruction of the source parameters (see, e.g., \citep{Contigiani:20aa,LIGO_lensing:21aa, Cheung:22aa}). 
Strongly lensed events moreover offer great opportunity to measure properties of the source population as well as of the Universe itself, such as the cosmological parameters (see, e.g., \citep{Sereno:10aa, Sereno:11aa, Liao:22aa} and references therein).\\
\indent In this work, we investigate galaxy-induced strong lensing on a population of EMRIs.
We work within the regime of geometrical optics since for LISA observational frequencies ($\sim$ mHz) we expect wave effects to become important only when considering diffusion from sub-galactic structures~\citep{PhysRevLett.80.1138,Cusin:2019rmt,Dalang:2021qhu,Caliskan:2023zqm} with masses $M \lesssim 10^{6}M_{\odot}$. 
In the context of LISA, we consider 12 EMRI population models, as described in \citep{Babak:17aa,Bonetti:2020jku}, and for each of them, we compute the 4-year detection rate of strongly-lensed EMRIs (LEMRIs). 
We then reconstruct the observed waveform of a LEMRI and we perform parameter estimation (PE) to recover parameters of both the source and the lens using a fully Bayesian approach.
Finally, assuming the LEMRI host galaxy can be uniquely identified, we discuss strategies that could be used to constrain cosmological parameters, in particular the Hubble constant $H_0$. \\
\indent Following \citep{Takahashi:11aa},  upon which our lensing rate calculations are based, we adopt a flat--$\Lambda$CDM cosmological model with matter density parameter $\Omega_{\rm m} = 0.274$, dark energy density parameter $\Omega_{\Lambda} = 0.726$ and Hubble constant $H_0 = 70.5 \,\text{km}\,\text{s}^{-1}\,\text{Mpc}^{-1}$. \\

\section{Gravitational lensing}
\label{sec:lensing}
In our work, we model galaxies as axisymmetric singular isothermal spheres (SIS), which are realistic enough for our purposes (see, e.g., \citep{Maoz:93, Mitchell:05aa}). Furthermore this approach allows us to derive simple analytical lensing formulae, that will be used later in Section \ref{sec:results}. \\
\indent In the following, we perform all the lensing calculations within the framework of geometrical optics, that is the scenario where the GW wavelength is smaller than the Schwarzschild radius of the lens. When operating within this regime, the primary lensing effects are signal magnification and generation of multiple images of the source, while diffraction effects are negligible (cf. Appendix \ref{appendix:geometric_regime}).

\subsection{The SIS model}
In the SIS model, the mass components of the galaxy behave like particles of an ideal gas, in thermal equilibrium, confined by their spherically symmetric gravitational potential \citep{Narayan:96aa}. The mass density of a SIS is
\begin{align}
    \rho=\frac{\varsigma^2}{2\pi G r^2},
\end{align}
with $\varsigma$ velocity dispersion of the lens and $r$ distance from its centre. The relation between the position of the source in the sky and its lensed image is given by the \textit{lens equation}, which reads \citep{Schneider:92aa}
\begin{align}
   \pmb{\beta}= \pmb{\theta} - \pmb{\alpha}.
    \label{eq:lens_eq}
\end{align}
In the above formula, $\pmb{\beta}$ is the angle between directions to the lens and to the source, $\pmb{\theta}$ is the angle between directions to the lens and to the image and $\pmb{\alpha}$ is the (scaled) \textit{deflection angle}, determined by the mass distribution of the lens projected along the line of sight \citep{Kochanek:06aa}. For a given SIS the deflection angle has a constant value
\begin{align}\label{eq:alpha_deflection}
    \alpha=4\pi^2 \left(\frac{\varsigma}{c}\right)^2\frac{d_{LS}}{d_{S}}\equiv \alpha_{0},
\end{align}   
with $d_{\rm LS}$ and $d_{\rm S}$ angular diameter distances \citep{hogg1999distance} between source-lens and source-observer respectively. In order to have the lensed signal splitted in multiple images, the following requirement needs to hold: $\beta < \alpha_{0}$.
\indent Defining the dimensionless variables $\pmb{y} \equiv\pmb{\beta}/\pmb{\alpha_0}$ and $\pmb{x}\equiv\pmb{\theta}/{\pmb{\alpha_0}}$, we can re-write Eq. \eqref{eq:lens_eq} in the following way 
\begin{align}
    \pmb{y}=\pmb{x}-\frac{\pmb{x}}{|\pmb{x}|}.
\end{align}
Thus, the criterion for multiple images becomes $y<1$ and the solutions of the lens equations are \cite{Cusin:21aa}
\begin{align}
    x_{+}=|\pmb{x_{+}}|=1+y,\,\,\,\,\, x_{-}=|\pmb{x_{-}}|=1-y.
\end{align}
The magnification that the two images will have due to the presence of the lens can be expressed in terms of $y$, \citep{Schneider:92aa}
\begin{align}\label{eq:magnification_factors}
    \mu_{+}=\frac{1}{y} + 1,\,\,\,\,\mu_{-}=\frac{1}{y} - 1,
\end{align}
and the difference between their arrival times at the detector is (see, e.g., \citep{Schneider:92aa,takahashi2003wave, Cusin:19aa})
\begin{align}
    \Delta t &= \frac{32\pi^2}{c}\left(\frac{\varsigma}{c}\right)^4\frac{d_{\rm L}d_{\rm LS}}{d_{\rm S}}(1+z_{\rm L})y \label{eq:delta_t_d_ls_stuff} \\
    & = 8 \frac{G M_{\rm Lz}}{c^3}y \label{eq:delta_t_M_lz}
\end{align}
with $M_{\rm Lz}=M_{\rm L}(1+z_{\rm L})$ being the redshifted lens mass.\\
\indent Within the geometrical optics limit, the lensed GW waveform reads ~\cite{bartelmann2001weak,takahashi2003wave}
\begin{equation}\label{eq:lensing_waveform}
    \hat{h}^{\rm L}(f) = (|\sqrt{\mu_{+}}| - i|\sqrt{\mu_{-}}|\exp[-2\pi i f \Delta t])\hat{h}(f),
\end{equation}
where $f$ is the gravitational frequency and $\hat{h}$ is the Fourier transform of the GW signal.

\subsection{Lensing prescription}
The number of lensed sources that we can observe up to redshift $z_{\rm max}$ with magnification higher than $\mu_{\rm min}$ is given by \citep{Cusin:21aa, Cusin:21ab, Toscani:23aa}
\begin{align}
{\mathcal{N}^{\rm obs}}=
&\int_{0}^{z_{\text{max}}}\text{d}z_{\rm S}\, \int_{\mu_{\text{min}}}^{\infty}\text{d}\mu \, p(\mu,z_{\rm S})\frac{\text{d}\mathcal{N}(\mu,z_{\rm S})}{\text{d}z_{\rm S}}\,.
  \label{eq:dn_obs}
\end{align}
The above formula is made of two main quantities: i) the magnification probability density function (PDF) for different values of source redshift, $p(\mu, z_{\rm S})$, and ii) the number of visible sources per bin of $z_{\rm S}$ in presence of magnification $\mu$, $\text{d}\mathcal{N}(\mu, z_{\rm S})/\text{d}z_{\rm S}$. In the following, we explain how we build these two quantities in the case of LEMRIs.

\subsubsection*{The magnification PDF}
We build the magnification PDF using the same methodology as the one already presented in \citep{Toscani:23aa}. Specifically, for cases where the source redshift $z_{\rm S}$ is less than 1, we compute $p(\mu, z_{\rm S})$ by distributing the lenses in accordance with the analytical model proposed by \citep{Cusin:21aa, Cusin:19aa}, which neglects the galaxy redshift evolution. This approach, validated by \citep{Cusin:21aa, Cusin:19aa}, yields accurate results at low redshifts, aligning well with more complex models, while allowing the use of analytical formulas. In addition, for situations where $z_{\rm S} \geq 1$, we incorporate the magnification PDF from \citep{Takahashi:11aa}, which was computed through high-resolution ray-tracing simulations reconstructing the path of light through inhomogeneous matter structures within the Universe. In particular, we interpolate their results for the redshift values we intend to investigate. For more technical details about the construction of the magnification PDF, we refer the reader to \citep{Toscani:23aa,Cusin:21aa,Cusin:19aa,Takahashi:11aa}.
\subsubsection*{The distribution of sources}
\indent Given a specific magnification factor $\mu$, the number of LEMRIs above the signal-to-noise ratio (SNR) detection threshold $\rho_{\rm th}$ per bin of source redshift $z_{\rm S}$ is given by  (cf. also Fig. \ref{fig:dn_dz_20_nobg} in Appendix \ref{appendix:emri_lemri})
\begin{align}
    \frac{\text{d}{\mathcal{N}}(\mu,z_{\rm S})}{\text{d}z_{\rm S}}=\int_{\rho_{\rm th}/\sqrt{\mu}}^{\infty}\text{d}\rho \, \frac{\text{d}\mathcal{N}}{\text{d}\rho \text{d}z_{\rm S}},
\end{align}
where $\text{d}\mathcal{N}/\text{d}\rho \text{d}z_{\rm S}$ is the number of sources per bin of SNR $\rho$ and bin of $z_{\rm S}$, that we compute for each of the 12 population models presented in \citep{Babak:17aa, Bonetti:2020jku}, which we similarly label M1 to M12.\\
\indent The main ingredients on which the EMRI models depend are the following \citep{Babak:17aa}:
\begin{enumerate}
    \item the MBH distribution - the MBH population is restricted to the range $10^{4}-10^{7}\,\text{M}_{\odot}$ and two different mass functions are considered. The first model \citep{Barausse:12aa} is derived from a self-consistent MBH evolutionary scenario and remains largely unaffected by changes in redshift. On the other hand, the second model \citep{Gair:10aa} is purely phenomenological and even if it is not derived from a self-consistent MBH evolutionary scenario is consistent with observational constraints. The latter model is employed for the analysis of M5 and M11. 
    \item CO mass - this is set equal to $10\text{M}_{\odot}$ across all population models, except for M4, which assumes a more massive CO of $30\text{M}_{\odot}$;
    \item ratio of plunges to inspirals - this is considered to be 0 for M7 and M12, 100 for M8 and M11 and 10 for all the other cases;
    \item MBH spin - three different distributions are considered: non-rotating MBH for M10 and M11, flat spin distribution over the interval (0,1) for M9 and highly rotating MBH with median value 0.98 for all the other populations;
    \item M-$\sigma$ relation - three different models are considered to describe the stellar distribution around the MBH and to predict the cusp regrowth after a MBH binary merger. A default distribution \citep{Gultekin:09aa} which estimates that the time for a cusp to reform is $t_{\rm cusp}\approx 6\,\text{Gyr}$, a more  pessimistic model \citep{Kormendy:13aa} (considered for M2) which gives $t_{\rm cusp}\approx 10\,\text{Gyr}$ and a more optimistic model \citep{Graham:13aa} (for M3) with $t_{\rm cusp}\approx 2\,\text{Gyr}$.
\end{enumerate}
\indent The interested reader can find more technical details and discussion about the different motivations behind each astrophysical scenario in \citep{Babak:17aa}.

\section{Data Analysis}
\label{sec:Data_Analysis}
\subsection{Preliminaries}\label{sec:preliminaries_data}
In our work we account for the LISA response~\cite{ armstrong1999time, vallisneri2005synthetic,katz2022assessing} acting on the two polarisations of the EMRI GW. The LISA detector will observe three data streams 
\begin{equation}\label{eq:data_stream}
    s^{(X)}(t) = h_{e}^{(X)}(t;\boldsymbol{\theta}^{E}_{\text{tr}}) + n^{(X)}(t) 
\end{equation}    
for first-generation TDI variables $X = \{A, E, T\}$. Here $h_{e}^{(X)}$ represents the exact responsed-GW signal with true EMRI source parameters $\boldsymbol{\theta}^{E}_{\text{tr}}$, and $n^{(X)}$ are noise realisations induced by instrumental perturbations and/or confusion noise. As a simplification, we assume that the noise in each channel, $n^{(X)}$, is an ergodic weakly-stationary Gaussian stochastic process with zero mean. Due to these assumptions, it is more convenient to work in the frequency domain. Our convention for the Fourier transform is
\begin{equation}
\hat{a}(f) = \mathcal{F}[h(t)] = \int_{0}^{\infty} a(t)\exp(-2\pi i f t)\, \text{d}t.
\end{equation}
As the noise is a stationary stochastic process, the noise is uncorrelated in the frequency domain. This results in a diagonal noise covariance matrix~\cite{wiener1930generalized,khintchine1934korrelationstheorie}
\begin{equation}
    \langle (\hat{n}^{(X)}(f))(\hat{n}^{(X)})^{\star}(f')\rangle = \frac{1}{2}\delta(f - f')S^{(X)}_{n}(f')
\end{equation} 
for $\langle \cdot \rangle$ an average ensemble over the data generating process, $\delta(f - f')$ Dirac delta function and $S^{(X)}_{n}(f)$ (one-sided) noise power-spectral-density (PSD) of the noise process $n^{(X)}(t)$ over each channel $X = \{A, E, T\}$.

In the results section~\ref{sec:Parameter_Estimation_Results} we will use the most realistic LISA orbits computed by the European Space Agency (ESA)~\cite{martens2021trajectory}. These orbits are carefully designed to account for the craft's gravitational interactions with significant celestial bodies within our solar system and the craft's fuel consumption, among other considerations. A result of this is that the arm-lengths are \emph{approximately} equal and \emph{approximately} constant. Although the orbit of the craft is more realistic, the variable arm-lengths introduces undesirable correlations between the noise components in the frequency domain~\cite{vallisneri2012non, vallisneri2005synthetic}. In our work we will neglect the effect of such correlations. This is reasonable since we are considering zero-noise injections and the impact of mis-modeling the noise process will impact the variability in recovered parameters (precision in a parameter estimation scheme). For a further review on this topic, see Refs.~\cite{burke2021extreme, edy2021issues, talbot2021inference}. \\
\indent The ESA based LISA-response alongside a GPU accelerated implementation is found in~\cite{katz2022assessing}. We use the code-base \texttt{lisa-on-gpu} to apply the response function, converting the source-frame EMRI polarisations to the solar-system barycenter (SSB) frame. For a further review on TDI, we refer the reader to Refs.~\cite{katz2021fast,armstrong1999time, vallisneri2005synthetic}. 

In our analysis, we use the usual noise-weighted inner product of the form 
\begin{equation}\label{eq:inn_prod}
(a|b)_{X} = \int_{0}^{\infty}\frac{a^{(X)}(f) (b^{(X)}(f))^{\star}}{S_{n}^{(X)}(f)} \, \text{d}f,
\end{equation}
and the so-called whittle-likelihood, which, for a known form of the PSD~\cite{whittle:1957}, reads
\begin{equation}\label{eq:whittle}
\log p(s|\boldsymbol{\theta}) \propto -\frac{1}{2}\sum_{X = \{A,E,T\}}(s - h_{m}(\boldsymbol{\theta})|s - h_{m}(\boldsymbol{\theta}))_{X}.
\end{equation}
The quantity $h_{m}(\boldsymbol{\theta})$ are our model templates, used to extract the underlying exact signal in the data stream (see Eq. \ref{eq:data_stream}).
\begin{figure*}
\includegraphics[width = \textwidth]{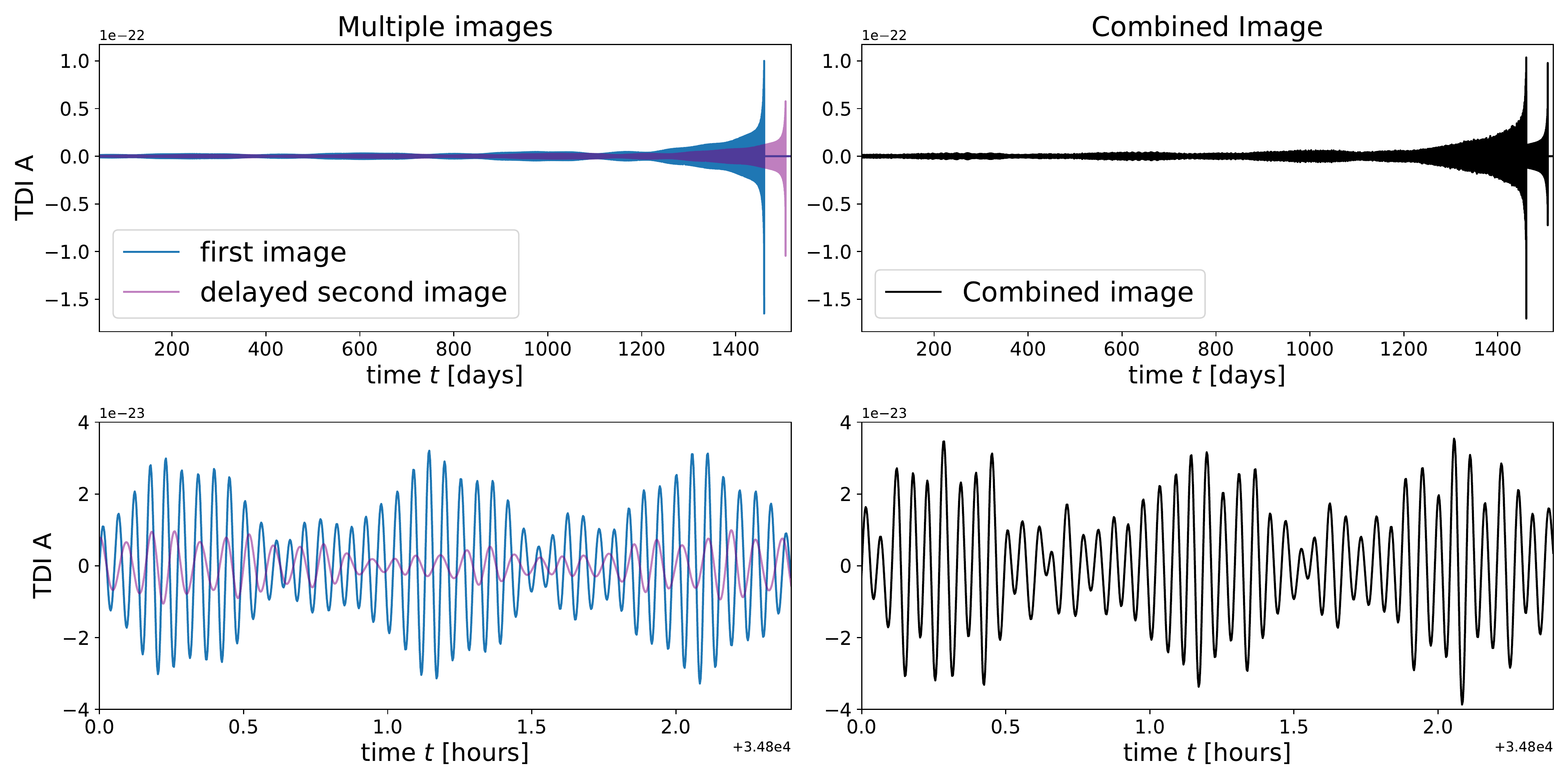}
\caption{\textbf{(Top left:)} TDI variable A plotted with respect to time (expressed in days), where the first-generation LISA response is applied. The blue and purple curves represent the magnified and de-magnified EMRI waveform in the time domain, respectively. \textbf{(Top right:)} sum of these two curves. \textbf{(Bottom line:)} same layout as above, but on a different time scale (hours).}
\label{fig:LEMRI_TD_plot}
\end{figure*}
In our analysis, we will assume that the noise properties are known and fixed. We use the latest \texttt{SciRDv1} noise curve~\citep{LISAsr:18aa} and account for the double-white-dwarf (DWD) confusion noise assuming LISA has been operating for 4 years~\cite{Cornish_2017}. We will not take into account the EMRI confusion background~\cite{Bonetti:2020jku}. The EMRI confusion background is the superposition of non-resolvable EMRI signals present within the data stream. Depending on the astrophysical assumptions of the generated EMRI background, the presence of such a background will raise the overall noise-floor or will be unnoticeable. If the most optimistic model is chosen (See M12 in Ref.~\cite{Babak:17aa}), then the SNR of the injected signal will lessen, increasing the uncertainty on measured parameters. The presence of the DWD background appears at almost the same frequency range of the EMRI background, and largely dominates. We thus believe that it is reasonable to exclude the EMRI background in the parameter estimation section~\ref{sec:Parameter_Estimation_Results}. 

We define the optimal matched-filtering signal-to-noise ratio (SNR) as the noise-weighted inner product with the injected signal and itself 
\begin{equation}\label{eq:SNR_AET}
\rho_{AET} = \left[\sum_{X = \{A,E,T\}}(h_{e}|h_{e})_{X}\right]^{1/2}\,,
\end{equation}
with $h_{e} := h_{e}(\boldsymbol{\theta}^{E})$. Eq. \eqref{eq:SNR_AET} measures the likelihood of detection as it measures the power of the signal to the root-mean-square average of noise fluctuations. For all EMRI signals generated throughout our work, we will always work with $\rho_{AET} \gtrsim 15$, since we believe this a suitable threshold for detection~\cite{Babak:17aa}.

\subsection{Parameter Estimation}\label{sec:PE}

Parameter estimation within gravitational wave astronomy is usually conducted using Bayesian methods. At the corner stone of Bayesian statistics lies Bayes' theorem, which states up to a constant of proportionality
\begin{equation}\label{eq:bayes_thm}
\log p(\boldsymbol{\theta}|s) \propto \log p(s|\boldsymbol{\theta}) + \log p(\boldsymbol{\theta}).
\end{equation}
Here $p(s|\boldsymbol{\theta})$ is the likelihood function, which is the probability of observing the data given a set of parameters of the source. In our analysis, we will use the whittle-likelihood (see Eq.~\ref{eq:whittle}). The second quantity on the right side, $p(\boldsymbol{\theta})$, represents our prior distributional beliefs of parameters $\boldsymbol{\theta}$ \textit{before} observing the data $d$. The missing term in Eq. \eqref{eq:bayes_thm}, the evidence $p(s) = \int p(s|\boldsymbol{\theta})p(\boldsymbol{\theta})\,\text{d}\boldsymbol{\theta}$, is a normalizing constant. It is a function only of the data, and we will ignore it in our analysis.  
The goal is to use algorithms to generate auto-correlated samples from the posterior density $p(\boldsymbol{\theta}|s)$, reflecting our beliefs on parameters \emph{after} observing the data stream $s$.

We will use Markov-Chain Monte-Carlo (MCMC) methods to generate samples from the posterior density. Our sampling algorithm of choice is the \texttt{emcee} algorithm~\cite{Foreman-Mackey:2013} (with default stretch proposal given by~\cite{2013PASP..125..306F}). In our simulations, we set 50 walkers (chains) to sample the log-posterior and set uninformative uniform priors on all parameters. Finally, we set the starting coordinates of our sampler $\boldsymbol{\theta}^{0} \approx \boldsymbol{\theta}_{\text{tr}}$. In our work, we are not performing a search, but instead wish to focus on constraints on parameters that govern the LEMRI system. During our simulations, we set the noise realizations in \eqref{eq:data_stream} to $n^{(X)}(t) = 0$. The noise process is suitably represented in the PSD for each channel, which features in the inner product \eqref{eq:inn_prod} and ultimately the likelihood \eqref{eq:whittle}. Incorporating noise realizations into our parameter estimation analysis will induce statistical fluctuations to the log-likelihood, enforcing the likelihood to no longer be maximized on the true parameters. The result of this is that posteriors are no longer centered on the true parameters. For our intents and purposes, it is unnecessary to include specific noise realizations. This is because the average statistical fluctuation enforcing a difference between the true parameters and recovered parameters is zero~\cite{cutler2007lisa}. Whether or not noise realizations are included, the conclusions of our results would not change. 

\subsection{LEMRI Waveform Model}\label{sec:waveform_model}
To generate our EMRI waveforms, we will use the latest state-of-the-art EMRI implementation \texttt{FastEMRIWaveforms} (\href{https://github.com/BlackHolePerturbationToolkit/FastEMRIWaveforms}{FEW})~\cite{katz2021fast,chua2021rapid} found in the Black Hole Perturbation Toolkit~\citep{BHPToolkit}. Our waveform model is the fully generic GPU accelerated 5PN AAK model that uses fifth-order post-Newtonian approximations to describe the trajectory of the CO driven by radiation reaction~\citep{chua2015improved, barack2004lisa}. The trajectories are valid at adiabatic order in the small mass-ratio $q = \mu/M \ll 1$ and neglect further contributions in the gravitational self-force at post-adiabatic order~\cite{Warburton:2021kwk,Wardell:2021fyy}. For a review on this topic, see~\cite{Pound:2021qin}. Our SSB responsed-EMRI waveform consists of a 14-dimensional parameter space, governed by $M$ and $\mu$ the primary/secondary objects redshifted masses, $a$ the dimensionless spin parameter, $p_{0}$ initial semi-latus rectum, $e_{0}$ initial eccentricity, $Y_{0} = \cos(\iota_{0})$ for $\iota_{0}$ initial inclination angle, $D_{\rm S}$ source luminosity distance in Gpc, $(\theta_{\rm S},\phi_{\rm S})$ angles describing the source location in the SSB frame, $(\theta_{\rm K},\phi_{\rm K})$ angles describing the orientation of the primary spin vector and $\{\Phi_{\rm r_0},\Phi_{\theta_0}, \Phi_{\phi_0}\}$ three initial fundamental frequencies intrinsic to the smaller body. \\
\indent The waveforms are generated with a sampling interval of 10 seconds and will be four years in length. The orbit (and thus waveform) terminates $\sim 0.1$ in semi-latus rectum away from the separatrix. The separatrix is defined as the final point in semi-latus rectum where stable orbits for the smaller companion object exist~\cite{stein2020location}. The reason for terminating the waveform at this point in semi-latus rectum is that the 5PN AAK model does not take into account the transition from inspiral to plunge. At the time of writing, there exist no fast-to-evaluate EMRI models that incorporate the transition from inspiral to plunge~\cite{ori2000transition,apte2019exciting,kesden2011transition,burke2020transition,compere2020transition}. The time-to-merger phase of an EMRI is hard to define, so instead the time-to-separatrix (or time to plunge $t_{\text{plunge}}$) is a more convenient parameter. Loosely speaking, the parameter $t_{\text{plunge}}$ is equivalent to the time to coalescence for comparable mass binaries. Unlike MBHs, EMRIs are not parametrized in terms of their \emph{time of coalescence} for two reasons: (1) because it is simply impractical and (2) the ``merger'' phase will likely be unobservable. 
This is due to a lack of SNR accumulated during those final few cycles during the plunging phase~\cite{ori2000transition}. To understand this, defining the small mass-ratio $q = \mu/M$, scaling arguments from~\cite{ori2000transition} show that the number of (azimuthal) orbits $\Phi_{\phi}$ scales 
\begin{align*}
\Phi_{\phi} &\sim \int \Omega_{\phi}\, \text{d}t \sim q^{-1}, \quad \text{inspiral,} \\
\Phi_{\phi} &\sim \int \Omega_{\phi}\, \text{d}t \sim q^{-3/5}, \quad \text{transition to plunge.}
\end{align*}
The number of orbits elapsed during the plunging phase is thus a mere fraction, approximately $q^{2/5}$ of the full number of orbits. Finally, since\footnote{From the two time-scale approach~\cite{Hinderer:2008dm} the number of orbits $N_{\text{orbits}}$ scales with the orbital time-scale, which is $\mathcal{O}(1)$ in comparison to the radiation reaction time-scale which is $\mathcal{O}(1/\eta)$. Approximating EMRIs as quasi-monochromatic sources, the SNR $\rho$ scales like $\sqrt{T_{\text{obs}}}$, which further scales with $\sqrt{N_{\text{obits}}}$.} the SNR $\rho \sim \sqrt{N_{\text{orbits}}}$, one can see that the SNR accumulation during the plunge phase is small in comparison to the inspiral phase. This is not the case for comparable high-mass binaries where instead the SNR is concentrated around the merger. As seen from the above scaling arguments, this is not the case for EMRIs. 

For computational convenience, EMRI waveforms are parametrized in terms of initial variables, $\{p_{0},e_{0},Y_{0}, \Phi_{\phi_{0}}, \Phi_{\theta_0}, \Phi_{r_{0}}\}$ that, together, dictate the initial time of observation.

The source is located at $z_{\rm S}=1.26$ since we expect to observe the largest number of LEMRIs between redshifts 1 and 2 (cf.~Figure~\ref{fig:observd_lemri}). With these chosen parameters, the optimal matched filtering SNR of the non-lensed source is $\rho^{\text{EMRI}}_{AET} \sim 22$, suitable for detection and parameter estimation. As for the lensing, we fix $y = 0.5$, resulting in magnification factors $\mu_{\pm} = \{3, 1\}$. For the lens, we choose a mass of $M_{\rm L} = 10^{11} \text{M}_{\odot}$, located at redshift $z_{\rm L} = 1$. With our choice of impact factor and redshifted lensed mass, the time-delay (see Eq. \ref{eq:delta_t_M_lz}) between the two images (signals) is $\Delta t \sim 45.6 \, \text{days}$.
Since $\Delta t \ll 1 \,$year, i.e., the orbital period of the craft, we can apply the response of LISA to the lensed waveform model (see Eq.~\ref{eq:lens_eq}) as a suitable approximation to the response of the instrument to the incoming double images sourced by the lens.

The SIS lens model results in two magnifications and a time-delay between the two images. The impact on the signal is a modified luminosity distance that controls the overall amplitude of the observed signal. We thus define two \emph{effective} luminosity distances, namely $D^{+} = D_{\rm S}/\sqrt{\mu_{+}}$ and $D^{-} = D_{\rm S}/\sqrt{\mu_{-}}$. The effective distance $D^{+} < D^{-}$, resulting in one image with higher SNR. The lensing parameters are thus $\boldsymbol{\theta}^{\rm L} = \{D^{+} := D_{\rm S}/\sqrt{\mu_{+}} = 5.13\ \text{Gpc}, D^{-} := D_{\rm S}/\sqrt{\mu_{-}} = 8.89 \ \text{Gpc}, \Delta t = 45.6 \ \text{days}\}$. With this configuration of parameters, the SNR of the LEMRI $\rho^{\text{LEMRI}}_{AET} \sim 45$. Approximately, this is an ehancement of the non-lensed EMRI $\rho_{AET}^{\text{EMRI}}$ by an amount consistent with the magnification factors $\sqrt{|\mu_{+}| + |\mu_{-}|} = 2$.

\indent We appreciate that $y = 0.5$ results in a magnification $\mu_{-} = 1$ giving rise to $D_{S} = D^{-}$. This was neither done on purpose, or a special case of the SIS lens. Our choice $y = 0.5$ was simply to ensure that the SNR of both images $\gtrsim 20$ to aid detection of each image separately. This is important, particularly in section \ref{sec:EMRI_detection_strat}. The full parameter space that will be sampled over is given by $\boldsymbol{\Theta} = \boldsymbol{\theta}^{\rm E}\cup \boldsymbol{\theta}^{\rm L}$. We generate the LEMRI using Eq.~\eqref{eq:lensing_waveform} and plot the first-generation TDI variable A as a function of time in Fig~\ref{fig:LEMRI_TD_plot}. The details presented in sections \ref{sec:PE} and \ref{sec:waveform_model} will be used specifically in the parameter estimation section \ref{sec:Parameter_Estimation_Results}.

For both waveform generation and application of the response, all computations are performed on a single NVIDIA A100 Tensor Core GPU using \texttt{cupy} as a drop-in replacement for \texttt{numpy} while utilising the GPU agnostic codes developed by \citep{katz2022assessing, katz2021fast}.

\begin{table*}
   \resizebox{1.5\columnwidth}{!}{
    \begin{tabular}{l|l|l|l|l}
\hline
\hline
& $\rho_{\rm th}=20$ & $\rho_{\rm th}=20$ & $\rho_{\rm th}=15$ & $\rho_{\rm th}=15$ \\
 &  w/o bg &w bg &   w/o bg & w bg \\ \hline
M1& 1.2 (482) & 0.6 (313 ) & 3.5 (1021)  & 2.1 (713)\\ \hline
M2 & 0.9 (358) & 0.5 (225) & 2.8 (804) & 1.7 (573) \\ \hline
M3 & 3.2 (1331) & 1.0 (689) & 7.9 (2638)& 3.6 (1436) \\ \hline
M4 & 7.8 (1846) & 6.6 (1587) & 12.3 (3156) & 11.4 (2827) \\ \hline
M5 & 0.2 (64) & 0.1 (63) &  0.3 (118) & 0.3 (116) \\ \hline
M6 & 1.9 (742) & 0.8 (435) & 5.0 (1508) & 2.7 (936)\\ \hline
M7 & 10.6 (4441) &0.7 (853)  & 31.8 (9544) & 2.8 (2091)\\ \hline
M8 & 0.1 (58) & 0.1 (51) &  0.3 (126) & 0.4 (120)\\ \hline
M9 & 0.7 (381) & 0.3 (231) &  2.2 (813) & 1.2 (555) \\ \hline
M10 &0.5 (346) & 0.3 (210) &  1.8 (753) & 1.0 (505) \\ \hline
M11 & 0.0  (1) & 0.0 (1)  &  0.0 (1)  & 0.0 (1)\\ \hline
M12 & 15.8 (6344) & 0.8 (1019) & 43.4 (13122) & 3.2 (2380) \\   
\hline 
\hline
\end{tabular}
   }
    \caption{
    4-year LEMRI (and EMRI) detection rates for LISA for all 12 EMRI population models. The first/last two columns consider an SNR threshold of 20/15 with and without taking into account the EMRI background confusion noise.
    }
    \label{ref:tab2} 
\end{table*}

\section{Results}
\label{sec:results}

\subsection{LEMRI event rates}
\label{sec:lemri_rates}

\indent In agreement with~\cite{babak2008mock,babak2009algorithm} we choose the following detection thresholds for LEMRIs, $\rho_{\rm th} \in \{15, 20\}$. As for the SNR calculation in this Section \ref{sec:lemri_rates}, in order to minimize the computational cost, we approximate the EMRI waveform with a inclination-polarization averaged simplified variate of the Analytical Kludge (AK) model~\citep{barack2004lisa}, as adopted also as in \cite{Bonetti:2020jku}. In particular, since we truncate the EMRI evolution at the Kerr innermost stable circular orbit~\cite{bardeen1972rotating}, we call it \textit{simplified AK Kerr (AKK) waveform}. Under this simplification, which follows the PN formalism from \cite{Peters1964}, the expression for the SNR calculation reads
\begin{equation}
(\mathrm{S} / \mathrm{N})_\text{AKK}^2=\int \frac{h_{\rm c}^2(f)}{f S_{\text {n}}(f)} \mathrm{d} \ln f .
\,,\end{equation}
Here, $h_{\rm c} = 2f \hat{h}(f)$ represents the total characteristic strain, which takes into account all harmonics, and $S_{\text {n}}(f)$ is the sky-averaged LISA sensitivity, specifically the \texttt{SciRDv1} sensitivity model \citep{LISAsr:18aa}, which also incorporates DWD confusion noise~\cite{Cornish_2017}, assuming a four-year observational period.

We remove sources with initial eccentricity $e_0 \gtrsim 0.9$ due to the limitations of the AKK model. 
Depending on the initial eccentricity, higher harmonics can be crucial components of $h_{\rm c}$, therefore we choose the number of harmonics $n$ carefully so that it's sufficient to keep the waveform accuracy. We take 
$n=20$ for $e_0<0.5$, 
$n=30$ for $e_0<0.8$, 
$n=40$ for $e_0<0.9$. Comparing the SNR calculated using the simplified AKK model $(\mathrm{S} / \mathrm{N})_\text{AKK}^2$ with the one calculated by 5PN AAK model $(\mathrm{S} / \mathrm{N})_\text{AAK}^2$, the ratio of $(\mathrm{S} / \mathrm{N})_\text{AKK}^2$ over $(\mathrm{S} / \mathrm{N})_\text{AAK}^2$ mostly distributed between 0.6 and 4.
Therefore in this section, using the simplified AKK waveform could lead to a slightly optimistic result in the detection rates of EMRI populations. 
We also take into account the EMRI confusion stochastic background computed by averaging over unresolved EMRIs according to various astrophysical models \cite{Bonetti:2020jku,Babak:17aa}.
For $\rho_{\rm th}=20$, we take the EMRI confusion backgrounds of the 12 population models from \citep{Bonetti:2020jku}, while for $\rho_{\rm th}=15$, in order to simplify calculations, we re-scale all the EMRI background PSDs by a factor of~$\sim1.18$, which is the EMRI background noise ratio between $\rho_{\rm th} \in [15,20]$ at frequency 3 mHz for population M1, as provided by~\citep{Bonetti:2020jku}.\\
\indent Table \ref{ref:tab2} presents LEMRI (and EMRI) detection rates for 4 years of LISA observations.
The detection rates are calculated for SNR thresholds of 15 and 20 and are provided for two different scenarios: with and without the inclusion of the EMRI confusion noise (for some extra information on the EMRI/LEMRI populations, see appendix \ref{appendix:emri_lemri}).\\
\indent Our results show that in 4 year LISA lifetime, the LEMRI detection rate ranges from 0 to $\sim 43$. In particular, excluding the most pessimistic population model M11, we can distinguish three cases: i) M4 always presents the highest percentage of observed LEMRIs (roughly $0.4\%$), in all the cases considered. This is mainly related to the fact that such a model extends to higher redshifts (up to $\sim 6$), where the contribution of lensing becomes more relevant (see, e.g., Discussion Section in \citep{Toscani:23aa}; ii) without the EMRI confusion noise, for both SNR thresholds, we have that the percentage of LEMRIs is around $0.2\%$ for all the other population models. In terms of absolute numbers, M7 and M12 present the highest number of detections (respectively, $\sim 11$ and $\sim 16$ for $\rho_{\rm th}=20$, while $\sim 32$ and $\sim 43$ for $\rho_{\rm th}=15$); iii) in presence of the EMRI background, we have that the percentage of observed LEMRIs remains roughly the same, except for M7 and M12, for which it drops to $\sim 0.1\%$, since these two populations are heavily affected by the background contribution. In terms of absolute numbers, in this case M4 is the more favorable scenario ($\sim 7$ for $\rho_{\rm th}=20$ and $\sim 11$ for $\rho_{\rm th}=15$).\\
\indent We want to stress that, in order to calculate the LEMRI rates, we require that the more magnified image exceeds the detectability threshold, thus ensuring that the superposition of the two images will be detectable by LISA.

\subsection{Parameter Estimation}\label{sec:Parameter_Estimation_Results}
In this section we will use the notions described in section \ref{sec:Data_Analysis} to characterize a LEMRI in a noiseless data stream. This section will be split into two parts: the first discusses a potential strategy to identify LEMRIs and the second will focus on estimating the LEMRI parameters ultimately leading to a constraint on the redshifted lens mass. 
\subsubsection{LEMRI detection strategy}\label{sec:EMRI_detection_strat}
We will begin with an injected LEMRI waveform with parameters described in section \ref{sec:waveform_model}, $\boldsymbol{\Theta} = \boldsymbol{\theta}^{E} \cup \boldsymbol{\theta}^{L}$, and attempt to characterize it with a single \emph{non-lensed} EMRI waveform model. In reality, during the search phase for EMRI signals, it is highly unlikely that a specific lensing model will be assumed prior to the first phase of searching for non-lensed EMRIs. Instead, it is more natural to search for a LEMRI first using a non-lensed EMRI template to gain intuition on what lensing model should describe the LEMRI within the data stream. In this section, we will describe a robust strategy that is able to extract a LEMRI within the data stream when working under the geometric optics limit. 

We wish to make clear here that we are \emph{not} performing a general search for the LEMRI within the data stream. We are assuming that suitable tools are in place that are able to efficiently search and characterize a single EMRI in the data stream. We are \emph{not} claiming a solution to this specific problem. The search problem, for EMRIs, has been solved in very simplified and ultimately unrealistic circumstances. Groups in the past~\cite{babak2009algorithm,gair2008improved, cornish2011detection,babak2008mock} submitted results to the Mock Lisa Data Challenges (MLDCs)~\cite{babak2008mock} using semi-analytical (Analytic Kludge~\cite{barack2004lisa}) EMRI waveforms buried in \emph{known} detector noise. We remind the reader here that the scope of this work is not to perform a search algorithm. Instead we will show that once a suitable search algorithm exists, it is possible to extract a LEMRI out the data stream without first assuming a lensing model. We refer the reader to Ref.~\cite{ye2023identification} for a recent study on the search problem of EMRI waveforms from eccentric Schwarzschild inspirals. 
\begin{figure*}
    \centering
    \includegraphics[width = \textwidth]{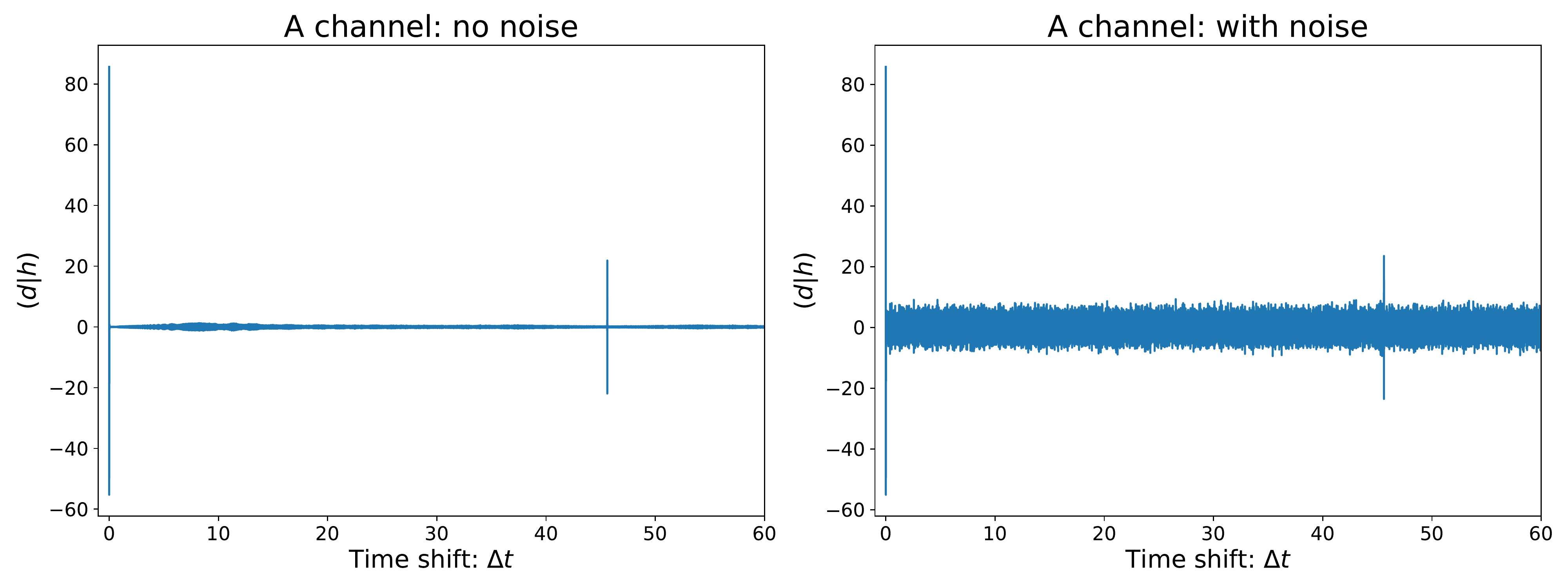}
    \caption{\textbf{Left plot:} Computation of Eq.~\eqref{eq:IFFT_delta_t} with a data stream composed entirely of signal. \textbf{Right plot:} is the same but a data stream with a signal embedded in Gaussian stationary noise. }
    \label{fig:IFFT_delta_t}
\end{figure*}

To begin, we inject a LEMRI with parameters stated in section \ref{sec:waveform_model} into three noise-free LISA data streams. We remind the reader that the lensing procedure produces two, \emph{largely orthogonal}, EMRI signals in the data stream (see Fig.~\ref{fig:LEMRI_TD_plot}). Performing parameter estimation on the LEMRI assuming a single EMRI model template results in the blue posterior distributions in Fig. \ref{fig:multiple_mismodel_PE} in Appendix. We report statistically insignificant biases on all parameters \emph{except} from the luminosity distance to the source. We recover a single magnified image (with respect to the non-lensed EMRI at $D_{\rm S}$) at an effective distance $D^{+} = D_{\rm \rm S}/\sqrt{\mu_{+}}$. Since the time-delay $\Delta t$ is (much) larger than the orbital-time scale of the smaller companion, the second image is orthogonal to the first image. This is the reason for the statistically insignificant biases on the intrinsic parameters. The second image can be thought of as a noise source with respect to the first detected image. 

At this point, we have identified a single EMRI (image) within the data stream, but there is still one image left to characterise. With the recovered parameters governing the single lensed image, it is possible to convolve this EMRI with the data stream to extract features of the data that are qualitatively similar to the first detected image. We discuss this procedure below. 

Recalling the Whittle-likelihood \eqref{eq:whittle} up to a constant of proportionality 
\begin{equation}
\log\mathcal{L} \propto (s|h_{\rm m}) - \frac{1}{2}(h_{\rm m}|h_{\rm m}),
\end{equation}
We wish to maximize the noise-weighted inner product $(d|h_{\rm m})$ by time-sliding the model templates evaluated at the best-fit parameters $h_{\rm m}(\boldsymbol{\theta}_{\text{bf}})$ across the data stream $d$. Time-shifts are trivial to encode in the Fourier domain 
\begin{equation}\label{eq:new_model_template_shift}
\hat{h}_{\rm m}(f;\boldsymbol{\theta},\Delta t) = \hat{h}_{\rm m}(f;\boldsymbol{\theta})\exp(-2\pi i f \Delta t).
\end{equation}
and thus we can re-write the inner product (Eq.~\ref{eq:inn_prod}) as
\begin{align}
(s|h) &= 4\int_{0}^{\infty}\left(\frac{\hat{s}(f)\hat{h}_{\rm m}^{\star}(f;\boldsymbol{\theta}_{\text{bf}})}{S_{n}(f)} \right)\exp(2\pi i f \Delta t)\, \text{d}f \, , \nonumber \\
& = \mathcal{F}^{-1}\left(4\cdot\frac{\hat{s}(f)\hat{h}^{\star}_{\rm m}(f;\boldsymbol{\theta}_{\text{bf}})}{S_{n}(f)} \right)(\Delta t)\, , \label{eq:IFFT_delta_t}
\end{align}
in other words, the inverse Fourier transform of the noise-weighted inner product with respect to time-shifts $\Delta t$. For $N = 2^{J}$ with $J \in \mathbb{Z}^{+}$, this is a $\mathcal{O}(N\log_{2}N)$ operation and is thus an inexpensive procedure. Eq.~\eqref{eq:IFFT_delta_t} is the \emph{matched-filter} statistic, used to cheaply detect patterns in the data stream that match a model template of choice.

To give an example of Eq.~\eqref{eq:IFFT_delta_t}, we consider the same LEMRI described above and consider two cases: with and without noise. The model template in \eqref{eq:IFFT_delta_t} will be a non-lensed EMRI with best-fit parameters $\boldsymbol{\theta}_{\text{bf}}^{E}$ given by the posterior means of the blue curves in Fig.~\ref{fig:multiple_mismodel_PE} in Appendix. For the noise and noiseless data stream, the results given by Eq.~\eqref{eq:IFFT_delta_t} are given in Fig.~\ref{fig:IFFT_delta_t}. Regardless of whether the signal is in the presence of stationary Gaussian noise, there are two clear maxima at $\Delta t = 0$ and $\Delta t \sim 45.6 \,$days. This implies that the first image has similar features to a second image present in the data stream. Due to the geometric optics limit, the specific frequency components of the signal are not changed. Only the amplitude is affected. As long as the signal is distinguishable from noise and wave-optics effects are negligible, this method will always return the correct time delay of the two images.

With starting coordinates at $\boldsymbol{\theta}_{\text{bf}}$ from the blue posterior with a time shift of $\Delta t \sim 45.6$ days, we perform another parameter estimation run assuming a non-lensed \emph{shifted} EMRI template \eqref{eq:new_model_template_shift}. Our results are given by the red posterior in Fig.~\ref{fig:multiple_mismodel_PE} in Appendix. Directly comparing with the blue posterior in the same Figure, we recover near-identical intrinsic/extrinsic parameters but recover a luminosity distance $D^{-} = D_{\rm S}/\sqrt{\mu_{-}}$ with respect to the true luminosity distance of the source $D_{\rm S}$. For both simulations, we notice that the same intrinsic parameters of the EMRI are recovered, with statistically insignificant biases on the extrinsic parameters except the luminosity distances. Due to our choice of $y=0.5$, we have that $\mu_{\pm} = \{3,1\}$ implying that $D^{-} = D_{\rm S}/\sqrt{\mu_{-}} = D_{\rm S}$ so it may seem that the result quoted here is unbiased. This is simply due to the choice of impact factor, had we chosen, say, $y = 0.8$ then the effective distances $D^{\pm} \neq D_{S}$ resulting in biased estimates of the luminosity distance. The take-home message here is that we have recovered two near-identical EMRI signals separated by $\Delta t$ with two different amplitudes. This is a strong indication that the source has been lensed and the lensing model is consistent with the SIS.  

From both Figs.~\ref{fig:multiple_mismodel_PE} and \ref{fig:IFFT_delta_t}, we have estimates of the \emph{two} (effective) luminosity distances to the source alongside a non-trivial time-delay between two images. If we now assume that the lensing model is consistent with the SIS, point-estimates for the lensing parameters can be easily computed. 
A point estimate of $y$ can be computed using
\begin{equation}\label{eq:y_and_effective_dist}
y = \frac{(D^{-})^2 - (D^{+})^2}{(D^{+})^2 + (D^{-})^2}\, .
\end{equation}
From Eq.~\eqref{eq:y_and_effective_dist}, our simulations indicate that $y \approx 0.509$. With the estimate $y$, one can use Eq.~\eqref{eq:magnification_factors} to compute the magnification factors $\mu_{\pm}$. This allows for a point estimate on the luminosity distance $D_{\rm S}$ of the source to be computed. Finally, from Eq. \eqref{eq:delta_t_M_lz}, a point estimate of the redshifted lensed mass can be returned. We remark here that since the results for the lensed mass and luminosity distance to the source come from two distinct simulations, correlations between the lensing parameters and EMRI parameters have been neglected. These correlations will be explored in section \ref{sec:PE_LEMRI_results} when a lensing model of the form \eqref{eq:lensing_waveform} is assumed when estimating the full parameter set $\boldsymbol{\Theta} = \boldsymbol{\theta}^{E} \cup \boldsymbol{\theta}^{L}$.

Above we have sketched out an algorithm on how point estimates can be returned for the lensing parameters. Now, we will argue why this detection strategy for a LEMRI is so robust. For EMRIs, the volume of the parameter space is huge~\cite{chua2022nonlocal}, so the chance of recovering two identical EMRIs, at the same position in the sky with identical intrinsic parameters but with different luminosity distances is near zero. The evolution of the individual frequencies of the EMRI signals are extremely sensitive to the intrinsic parameters of the system. The two images, with only differences being the time-delay and the magnifications, should never admit biased parameter simulations when searching with a single EMRI template. This is because the two images will be orthogonal assuming that the time-delay is greater than the orbital time-scale of the smaller compact object. 

The deflection angle $\alpha$ (Eq. \eqref{eq:alpha_deflection}) will also be small. Since $\alpha$ is small and $\Delta t \ll 1\,$ year, the orbital time-scale of LISA, there should be little to no bias on the sky-location parameters. There may be small biases on the initial frequencies $\Phi_{\phi_{0}}, \Phi_{\theta_{0}}$ and $\Phi_{r_{0}}$, likely arising from the geometry of the instrument changing over the scale of the time-shift. Even if the transition to inspiral to plunge was incorporated in our model, the plunging phase would be largely unresolvable. Inclusion of the transition from inspiral to plunge would neither help or hinder our detection strategy. We thus conclude that, within the geometric optics regime, the only noticeable biases would be on the luminosity distances to the source. This, alongside maximizing over all time-shifts $\Delta t$, would give evidence for a particular lensing model to assume when re-re-analysing the data stream. In our case it was the SIS model, however, multiple images could be returned implying more complex lensing models could be assumed such as a singular isothermal ellipsoid \citep[see, e.g.][]{Kormann:94aa}. 

\subsubsection{Constraining LEMRI parameters}\label{sec:PE_LEMRI_results}

In this Section, we will assume the SIS lensing model and constrain \emph{both} the EMRI parameters $\boldsymbol{\theta}^{E}$ and the parameters specific to the lens $\boldsymbol{\theta}^{L} = \{D^{+}, D^{-}, \Delta t\}$. With the constraints from the lensing parameters, we will be able to estimate $y$, $M_{Lz}$ and the ratio of angular diameter distances $d_{\rm L}d_{\rm LS}/d_{\rm S}$. The constraint on this ratio will be useful for studies within cosmology that we present in section \ref{sec:cosmology}.

We can rewrite equation \eqref{eq:lensing_waveform} 
\begin{equation}\label{eq:mcmc_model_template_lens}
    \hat{h}^{\rm L}(f;\boldsymbol{\Theta}) = \left(\frac{1}{D^{+}} - \frac{i}{D^{-}}e^{-2\pi i f \Delta t}\right)\hat{h}(f;\boldsymbol{\theta}^{E}\backslash \{D_{S}\}).
\end{equation}
To aid the parameter estimation scheme, we can use the results from the previous section given in Fig.~\ref{fig:multiple_mismodel_PE} in Appendix. Using the \texttt{emcee} sampling algorithm, we can set the starting coordinates to be the posterior means of the recovered parameters given by the blue (image 1) and red (image 2) posteriors. Similarly, we can set the starting coordinate for the time-delay to be $\Delta t \approx 45.6\,$ days, in accordance with Fig.~\ref{fig:IFFT_delta_t}. In our simulation, we will sample over the full EMRI parameter space $\boldsymbol{\theta}^{E}$ and the lensing parameters $\boldsymbol{\theta}^{L} = \{D^{+}, D^{-}, \Delta t\}$. The goal here is to understand how well one can constrain the lensing parameters and whether correlations between the two parameter sets $\boldsymbol{\theta}^{E}$ and $\boldsymbol{\theta}^{L}$ exist.

We inject a LEMRI into noiseless data and recover with a lensing model template given by Eq.~\eqref{eq:lensing_waveform}. Our results are given in Fig.~\ref{fig:LEMRI_PE_Plot} in Appendix. With samples for the effective distances, we can use both \eqref{eq:y_and_effective_dist} and \eqref{eq:delta_t_M_lz} to determine samples for $M_{Lz}$. Since the EMRI and lensing parameter sets are largely orthogonal, we see that the only correlated parameter set is between the luminosity distance and the redshifted mass of the lens $M_{\text{Lz}}$. Defining the error in the sky location, $\Delta \Omega$ as in~\cite{cutler1998angular},
\begin{equation}
 \Delta \Omega = 2\pi (\Delta \sin(\theta_{S})\Delta{\phi}_{S} - \text{Cov}[\sin(\theta_{S}),\phi_{S}])
\end{equation}
we can constrain the sky-position of the LEMRI by $\Delta \Omega \sim 11\,\text{deg}^2$. The relative errors on the redshifted mass of the lens and source impact parameter are $\Delta M_{\rm Lz}/M_{\rm Lz}  \lesssim 4\%$ and $\Delta y/y \lesssim 4\%$. This is consistent with Fisher matrix estimates given by Eq.~(34) in~\cite{takahashi2003wave}. 

The LEMRI parameters are distributed as Gaussians with tight constraints, as expected in the unlensed case \cite{Babak:17aa, katz2021fast}. In both  scenarios, the orientation of the spin vector $(\theta_{\rm K},\phi_{\rm K})$ shows degeneracy, consistent with previous analyses~\cite{babak2009algorithm, babak2008mock}, while the distance parameter is poorly constrained only for the secondary image, likely due to the low SNR of the injected signal ($\rho_{\rm EMRI} \sim 20$). We believe that the $\theta_{K},\phi_{K}$ degeneracy originates from using semi-relativistic waveform models with PN driven inspirals. Such effects have not been observed in fully relativistic waveforms with accurate adiabatic flux driven inspirals originating from the self force~\cite{Burke:2023lno}. 

As a comparison, we have performed another simulation where we inject a non-lensed EMRI into noiseless data with an identical parameter set to the LEMRI. The precision measurements degrade on parmeters by an amount consistent with the loss of SNR due to the lack of magnification incurred by the lens.  

Our findings suggest that LISA can successfully identify LEMRIs even in the case when the original unlensed signal is close to the SNR threshold ($\rho_{\rm th} \sim 20$). The parameter constraints improve in accordance with the factor $\sim \sqrt{|\mu_{+}| + |\mu_{-}|}$, as expected. In addition, we investigate a more optimistic LEMRI with the same parameters as the previous injection, apart from $z_{\rm S} \sim 0.5$ and $z_{\rm L} \sim 0.3$. Over all TDI channels, we find that $\rho_{\rm LEMRI} \sim 140$. Consequently, the relative errors on the redshifted lens mass and the luminosity distance of the source improve to the $\sim 1\%$ level. In addition to this, for such a luminous LEMRI, the sky position could be determined within a better accuracy  ($\Delta \Omega \sim 3 \,\text{deg}^2$), and also the posteriors for spin orientation $(\theta_{\rm K},\phi_{\rm K})$ resemble more Gaussian behavior.

\subsection{\label{sec:cosmology} Cosmology}

\subsubsection{Constraining $H_{0}$}
\label{sub:H0}
\begin{figure}
    \includegraphics[width = 0.45\textwidth]{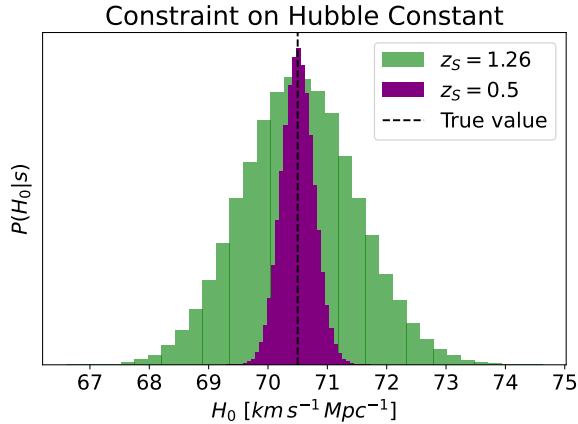}
    \caption{Posterior distributions on the Hubble constant for a LEMRI located at $z_{\rm S}=0.5$ (purple) and at $z_{\rm S}=1.26$ (green), assuming unique identification of the host galaxy and a flat $\Lambda$CDM model with $\Omega_m$ fixed to its fiducial value. The injected value of $H_0$ is represented by the black dashed vertical line.}
    \label{fig:Hubble_Constant_Constraint}
\end{figure}

Given its reduced sky-localization and the constraints we can infer on its parameters as outlined above, the lens and the host galaxy, which also undergoes strong lensing, may be identified through dedicated deep-field follow-up EM surveys.
By leveraging on EM follow-up strategies developed for Earth-based GW detectors~\cite{Hannuksela:2020xor,Wempe:2022zlk}, and by noting that LEMRIs will convey additional information on the host galaxy luminosity since the MBH mass is tightly constrained, we can optimistically expect that lens and host galaxy can be spotted for a relevant fraction of observed LEMRIs. This strategy can be applied also to the case where only two lensed images are produced \citep{Wempe:2022zlk}, as considered in this work.\\
\indent Working within a flat-$\Lambda$CDM model, and assuming a spectroscopic redshift measurement of the host galaxy (which implies a redshift uncertainties irrelevantly small for our estimates), we can then constrain $H_{0}$ if we fix $\Omega_{\rm m} = 0.274$.
Note that we can only constrain a one-parameter cosmological model with one single LEMRI event because we obtain one single point in the distance-redshift space which would yield strong degeneracy for a multi-parameter model.
In particular, we consider two estimates of $H_0$ derived from two separately investigated LEMRIs at $z_{\rm S} = \{1.26, 0.5\}$, respectively with lenses at $z_{\rm L} = \{1, 0.3\}$, corresponding to the two sets of injected values we considered above.\\ 
\indent From parameter estimation simulations, ignoring systematic uncertainties due to weak lensing and peculiar velocities, we find that the luminosity distances can be constrained to within $\Delta D_{\rm S} / D_{\rm S} \lesssim 2\%$ and $\Delta D_{\rm S} / D_{\rm S} \lesssim 1\%$, respectively (1$\sigma$ C.I.). 
From the distance-redshift relation we obtain Fig.~\ref{fig:Hubble_Constant_Constraint}, which shows the posterior distributions $p(H_{0}|s)$ for each case. Our results demonstrate that $H_{0}$ can be constrained to within a relative error of around $\sim 0.4\%$ for the LEMRI at $z_{\rm S} = 0.5$, while the relative error increases to $\sim 1\%$ at $z_{\rm S} = 1.26$.
Such measurements are comparable to, or even more precise of, other cosmological forecasts with LISA \cite{Tamanini:2016zlh,LISACosmologyWorkingGroup:2019mwx,Laghi:2021pqk,LISACosmologyWorkingGroup:2022jok} and they would definitely be enough accurate to provide a solution to the Hubble tension \cite{Feeney:2018mkj}.
Furthermore, they are more realistic than similar cosmological estimates with strongly-lensed LISA MBHBs \cite{Sereno:10aa,Sereno:11aa} since identification of the EMRI host galaxy may be easier given the lower expected redshift of the source and better sky-localization, directly providing less potential hosts in the cosmological localization volume of the source.
On the other hand, in our analysis we ignored possible selection effects due to weak lensing, peculiar motion, waveform and calibration uncertainties which may well provide additional systematic errors at the percent level or more if not properly accounted for (see e.g.~\cite{Tamanini:2019usx,Purrer:2019jcp,Cusin:2020ezb,Huang:2022rdg,Hu:2022rjq}).

\subsubsection{Constraining $H_{0}$ and $\Omega_{\rm m}$}
We also want to investigate a different approach to constrain both $H_0$ and $\Omega_{\rm m}$, assuming further knowledge of the redshift and velocity dispersion $\varsigma$ of the galaxy-lens.
Such a scenario could materialize only if the lens can be identified with EM observations and its morphology is well characterized (see e.g.~\cite{Hannuksela:2020xor,Wempe:2022zlk}).

With knowledge of both $z_{\rm L}$ and $\varsigma$ (both of which we assume with zero uncertainty for our analysis here, neglecting peculiar velocity effects), and with our estimates for $y$ and $\Delta t$, it is possible to constrain the ratio of the angular diameter distances $d_{\rm L}d_{\rm LS}/d_{\rm S}$.
Knowledge of this ratio of quantities can then lead to an estimate to the luminosity distance to the lens. The angular diameter distance to an object is related to the luminosity distance via $d_{\rm L} = (1 + z_{\rm L})^2 D_{\rm L}$. Similarly, assuming a flat cosmology with $\Omega_{\rm k} = 0$, the angular diameter distance between the lens and source $d_{\rm LS}$ is given by~\cite{hogg1999distance}
\begin{equation}
d_{\rm LS} = \frac{1}{1 + z_{\rm S}}\left(\frac{d_{\rm S}}{1 + z_{\rm S}} - \frac{d_{\rm L}}{1 + z_{\rm L}}\right).
\end{equation}
With this equation, one can write down a quadratic equation for the luminosity distance $D_{\rm L}$
\begin{equation}
\frac{1 + z_{\rm S}}{D_{\rm S}(1 + z_{\rm L})^3}D^2_{\rm L} - \frac{1}{(1 + z_{\rm L})^2}D_{\rm L} + \frac{d_{\rm L}d_{\rm LS}}{d_{\rm S}} = 0
\end{equation}
with (positive) solution
\begin{equation}\label{eq:dist_lens}
D_{\rm L} = \frac{D_{\rm S}(1 + z_{\rm L})}{2(1+z_{\rm S})}\left(1 + \sqrt{1 - 4 \frac{d_{\rm L}d_{\rm LS}(1 + z_{\rm S})(1+z_{\rm L})}{d_{\rm S}D_{\rm S}}}\right).
\end{equation}

\begin{figure}
    \includegraphics[width = 0.45\textwidth]{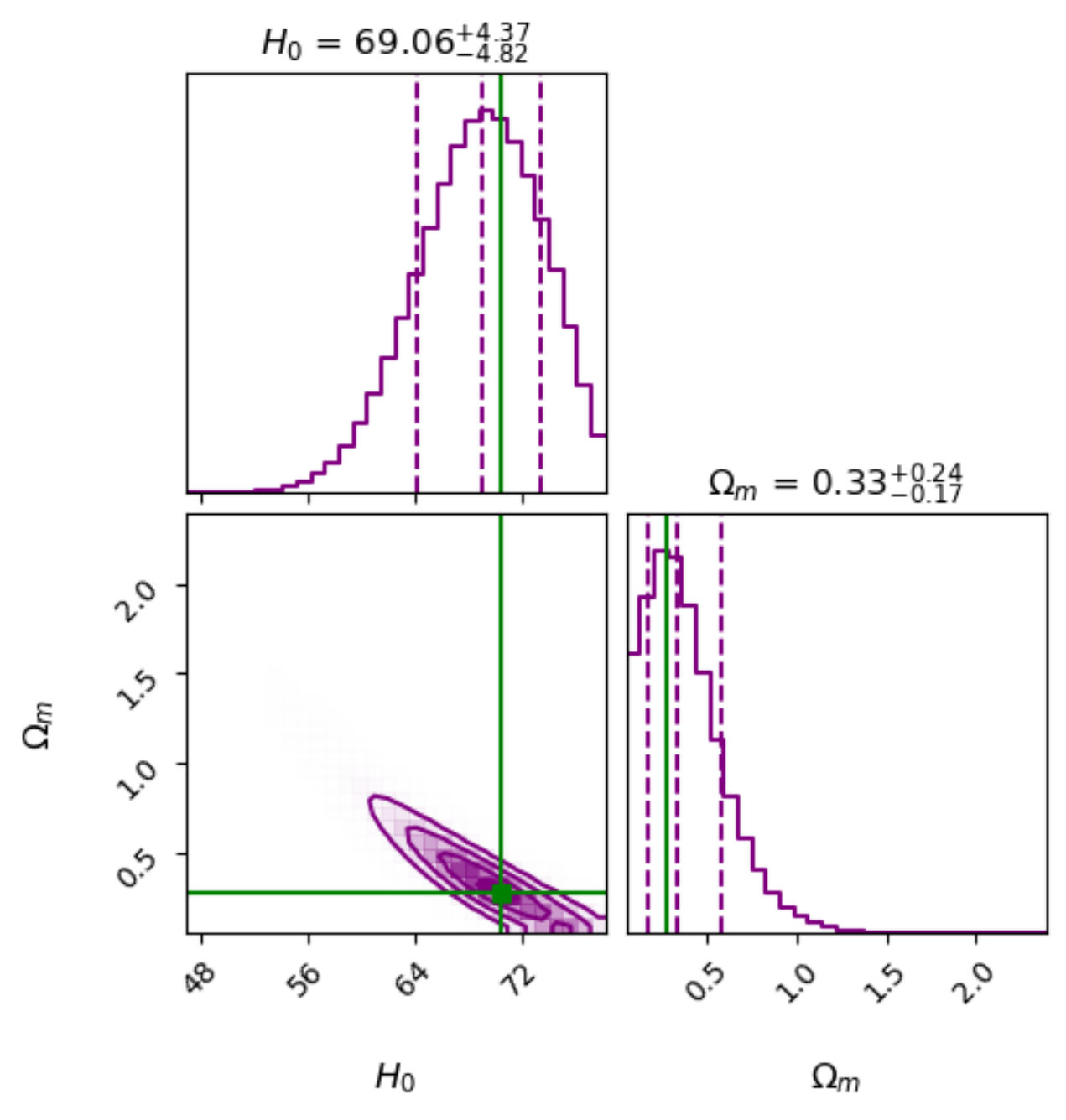}
    \caption{
    Simultaneous constraints on both $H_0$ and $\Omega_m$ assuming precise EM identification and characterization of the LEMRI lens.
    The purple posteriors represent one/two dimensional marginalizations of $p(H_{0},\Omega_{\rm m}|s, z_{L} = 0.3, z_{S} = 0.5)$. The green lines indicate the injected values $\Omega_{\rm m} = 0.274$ and $H_0 = 70.5 \,\text{km}\,\text{s}^{-1}\,\text{Mpc}^{-1}$.}
    \label{fig:Hubble_Constant_Omega_Constraint}
\end{figure}

The luminosity distance to the lens is then calculated from Eq.~\eqref{eq:dist_lens}. From the distance-redshift relation, the estimates of $(z_{\rm S},D_{\rm S})$ and $(z_{\rm L},D_{\rm L})$ could allow for constraints on the two cosmological parameters $(H_{0},\Omega_{\rm m})$ simultaneously.
However, with significant uncertainties on $\varsigma$ and the difficulty of properly measuring $z_{\rm L}$, these estimates could be seen as largely optimistic.
We remark here that neither cosmological parameter is well-constrained using the weaker source at $z_{\rm S}=1.26$.
For a loud EMRI with $z_{\rm S} = 0.5$ instead, it is possible to constrain the luminosity distance to the lens $\Delta D_{\rm L} /D_{\rm L}  \lesssim 3\%$ (1$\sigma$ C.I.) giving rise to the simultaneous constraints on $H_0$ and $\Omega_{\rm m}$ presented in Fig.~\ref{fig:Hubble_Constant_Omega_Constraint}.
These results indicate that one can constrain $H_{0}$ to within a relative error of $\Delta H_{0}/H_{0} \lesssim 7\%$  whilst we weakly constrain $\Delta \Omega_{\rm m}/\Omega_{\rm m} \lesssim 60\%$ (1$\sigma$ C.I.).
These statements are statistically expected to scale with the number of LEMRI observations as $\sim N_{\text{obs}}^{-1/2}$.
We stress that, contrary to the $H_0$ estimates presented in Sec.~\ref{sub:H0}, these results do not assume prior knowledge on $\Omega_{\rm m}$.
They moreover strongly depend on our SIS modeling of the lens, although the EM follow-up of the lens may provide sufficient morphological information to understand which lensing model may better describe the LEMRI observations and thus provide unbiased estimates.

\section{\label{sec:level5} Conclusion}
In this work, we show that LEMRIs are new potential targets for LISA and we outline a technique to extract this signal from the data-stream. The identification of multiple images of the same source would allow us to understand what is the best model that describes the lens, giving valuable information on the nature of this object. Furthermore, the unique identification of the LEMRI host galaxy would yield an unprecedented opportunity to understand in depth the nature of the source and the lens, as well as to perform precise cosmological measurements.
Our results show that LEMRIs constitute a new target GW signal for LISA, whose observation will enable interesting astrophysical and cosmological insights. Further work is needed to fully characterize their detectability and scientific potential.

\section*{Aknowledgments}
M.~Toscani expresses gratitude to Giulia Cusin for her guidance in understanding the complexities of lensing. The authors thank Jose Maria Ezquiaga for useful discussions about gravitational lensing and Matteo Bonetti, Alberto Sesana and Stas Babak for discussions and data on EMRI populations and EMRI background.
O.~Burke expresses his gratitude for many useful discussions with Sylvain Marsat, Lorenzo Speri, Mesut Çalışkan and thanks Michael Katz for writing an extremely user-friendly code base. Both O.~Burke and M.~Toscani have also enjoyed the vocal support from Shakira whilst waiting for our simulations to end.
O.~Burke, M.~Toscani, C.~Liu and N.~Tamanini acknowledge support from the French space agency CNES in the framework of LISA.
This project has received financial support from the CNRS through the MITI interdisciplinary programs. C.~Liu is supported by the China Scholarship Council (CSC) and the National Natural Science Foundation of China (11991053, 11975027).
This work has made use of \texttt{emcee}, \texttt{scipy}, \texttt{numpy},  \texttt{matplotlib}, \texttt{corner}, \texttt{cupy}, \texttt{astropy}, \texttt{multiprocess} and \texttt{lisatools}~\citep{2020SciPy-NMeth, harris2020array, 2013PASP..125..306F,Hunter:2007, corner, cupy_learningsys2017, astropy:2022, mp, KatzTools}.  This work uses the \href{https://bhptoolkit.org}{Black Hole Perturbation Toolkit}.\\

\section*{Author Contributions}
MT: Conceptualization, Formal Analysis: lensing, Investigation, Methodology, Software: development of the code for the LEMRI rates, Visualization, Writing -- original draft, Writing - review and editing.
\newline \indent OB: Conceptualization, Formal analysis: Parameter estimation \& Data analysis, Investigation, Methodology, Software: MCMC codes, waveform generation, Visualization: Corner plots, Writing -- original draft, Writing -- Review and Editing.
\newline \indent CL: Conceptualization, Formal Analysis: (unlensed) population, Investigation, Visualization, Writing - original draft, Writing - review and editing.
\newline \indent NZ: Investigation, Visualization, Writing -- editing.
\newline \indent NT: Conceptualization, funding acquisition, project administration, resources, supervision, validation, writing - review and editing.
\newline \indent FP: Resources,  Writing - review and editing.

\appendix
\section{Geometric optics regime}
\label{appendix:geometric_regime}

The validity of Eq.~\eqref{eq:lensing_waveform} is determined by the magnitude of 
\begin{equation}
\omega = \left(\frac{8\pi G}{c^3}\right) \cdot f M_{\text{L}z} \approx \left(\frac{M_{\text{L}z}}{10^{7}M_{\odot}}\right)\left(\frac{f}{\text{mHz}}\right)\,,
\end{equation}
where $\omega$ determines the oscillatory nature of the general lensing amplification factor in Eq.~(19) of~\cite{takahashi2003wave}. For the EMRI source considered in this work, we take $M_{\text{L}} = 10^{11}M_{\odot}$ and $z_{\text{L}} = 1$ with an observation period of $T_{\text{obs}} = 4$ years, with cadence $\sim 10$ seconds. From a signal processing viewpoint, the resolvable frequencies span $f\in [0, 0.05]\,$Hz spaced equally with $\Delta f = 1/T_{\text{obs}} \sim 10^{-10}\,$Hz. However, due to the presence of LISA instrumental noise, we are insensitive to frequencies $f \not\in [10^{-5},10^{-1}]\,$Hz. For the source considered in this work, we find that $1 \ll \omega \in (10^{2}, 5\cdot 10^{5})$ thus validating the geometric optics simplification used throughout our work. Effects due to wave optics will always be present, but largely insignificant since they are subdominant contributions to the overall lensing effect. If, on the other hand, we were sensitive to lower frequency sources $f \sim \mu$Hz and/or had a less massive lens mass (such as a MBH) $M_{\text{L} z} \sim 10^{7}M_{\odot}$, then we must consider potential wave-optics effects. For more details on gravitational wave science from a wave-optics perspective, see~\cite{takahashi2003wave, ccaliotacskan2023observability, ezquiaga2021phase}.

\section{Additional information on the EMRI/LEMRI population properties}
\label{appendix:emri_lemri}
\begin{figure}
    \includegraphics[width = 0.45\textwidth]{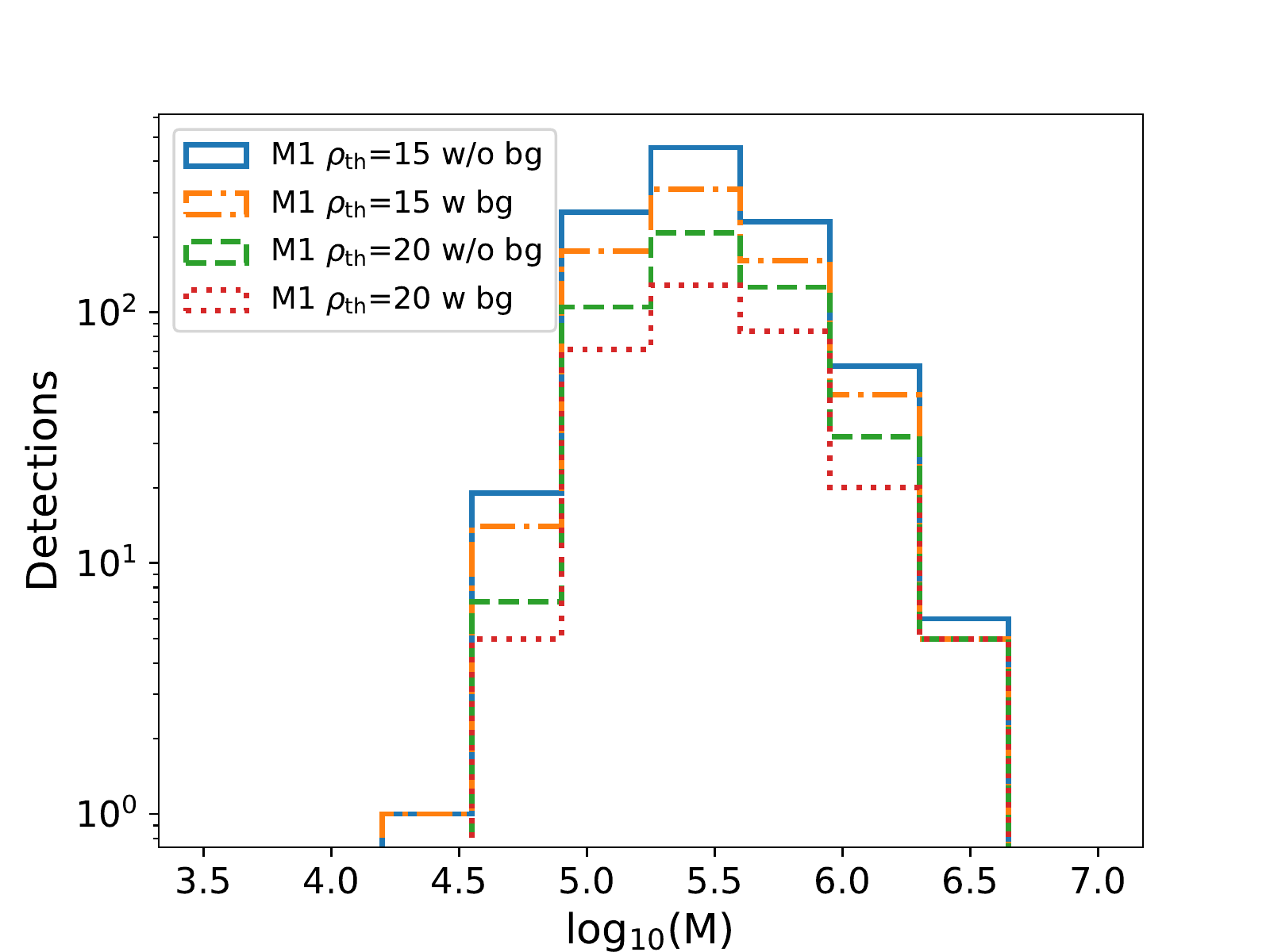}
    \includegraphics[width = 0.45\textwidth]{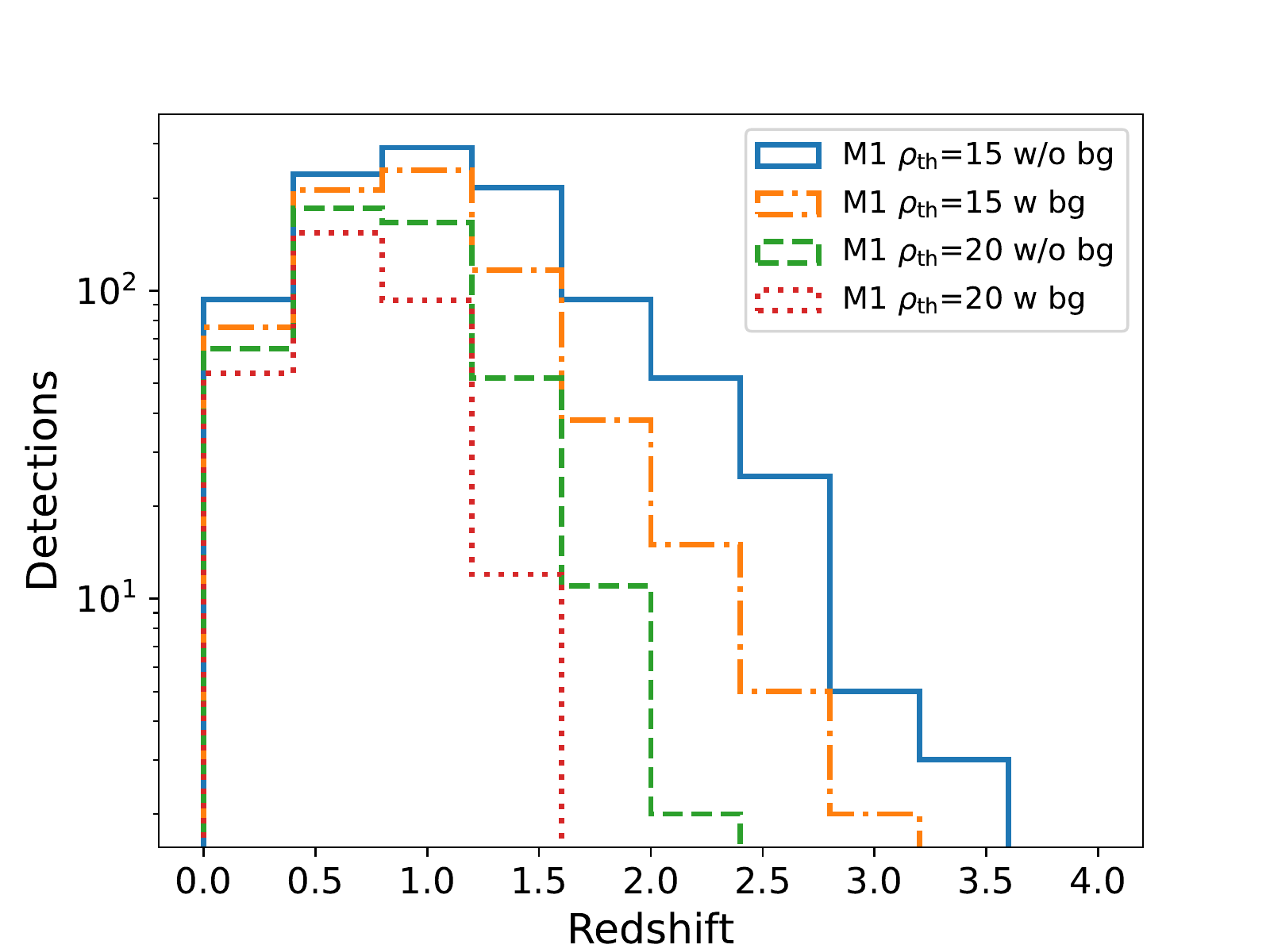}
    \caption{The MBH mass (upper panel) and redshift (lower panel) distributions of the detected EMRIs for population model M1 after 4 years of LISA observation.}
    \label{fig:M1}
\end{figure}

\begin{figure}
    \includegraphics[width = 0.45\textwidth]{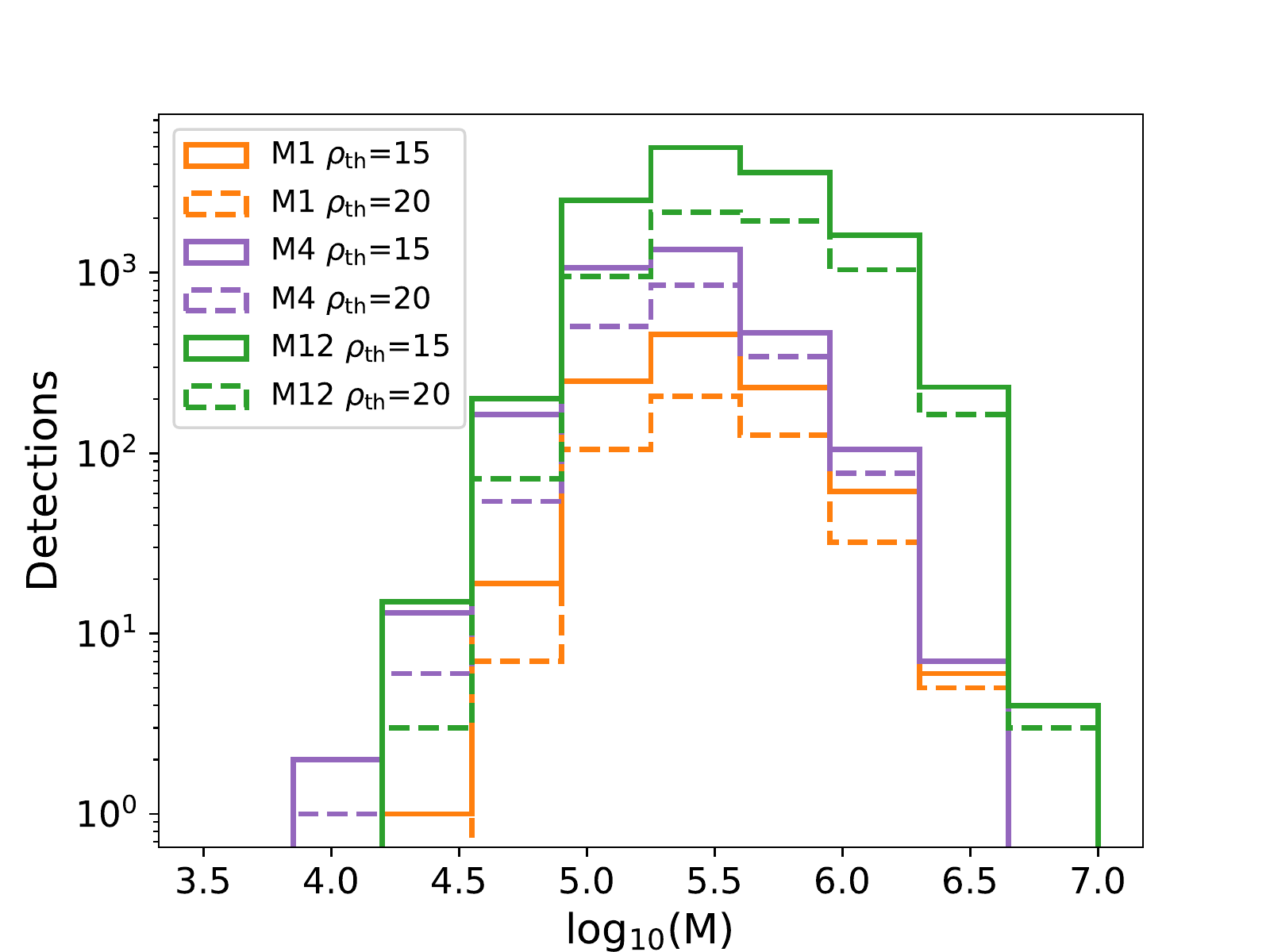}
    \includegraphics[width = 0.45\textwidth]{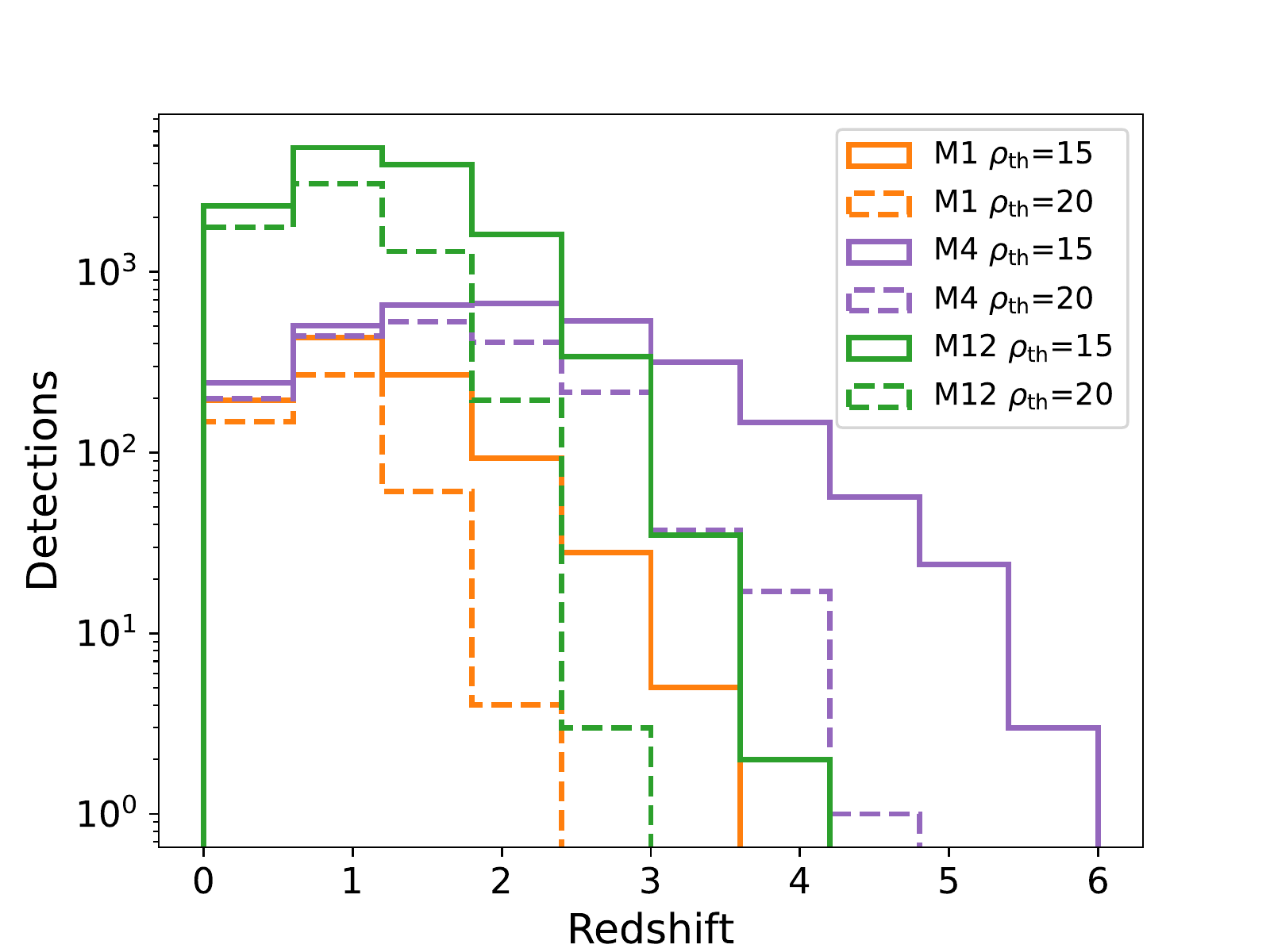}
    \caption{The MBH mass (upper panel) and redshift (lower panel) distributions of the detected EMRIs for population model M1, M4 and M12 after 4 years of LISA observation without EMRI background confusion noise. The solid (dashed) lines show the distributions at SNR threshold 15 (20).}\label{fig:M1M4M12}
\end{figure}

\begin{figure*}
    \centering
    \includegraphics[width = \textwidth]{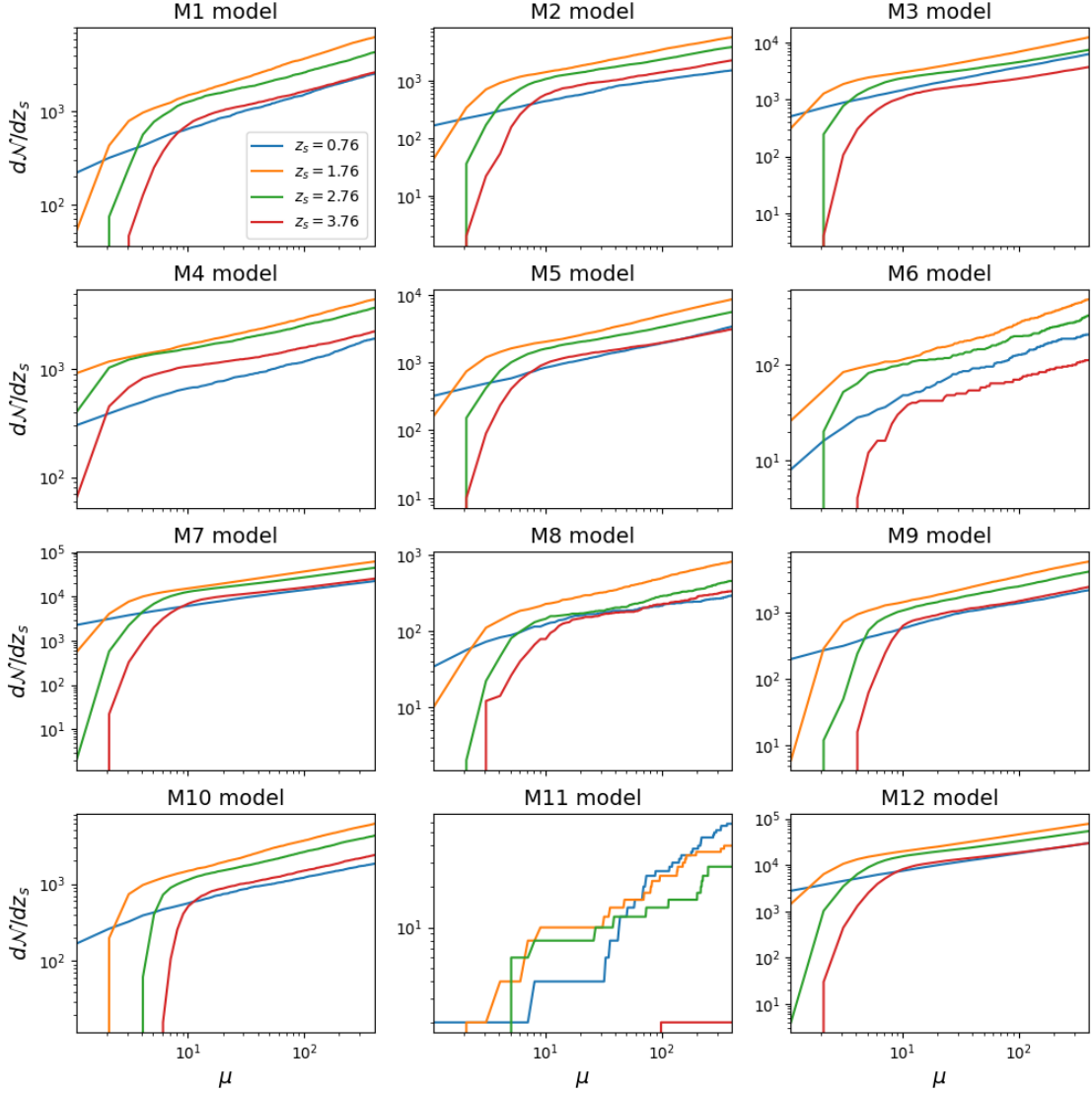}
    \caption{Number of LEMRIs per redshift bin assuming $\rho_{\rm th}=20$ and no EMRI confusion noise. We show the curves for selected redshift bins centered on the following values: $z_{\rm s}=0.76,1.76,2.76,3.76$. }
    \label{fig:dn_dz_20_nobg}
\end{figure*}

\begin{figure}
    \centering
    \includegraphics[width = 0.4\textwidth]{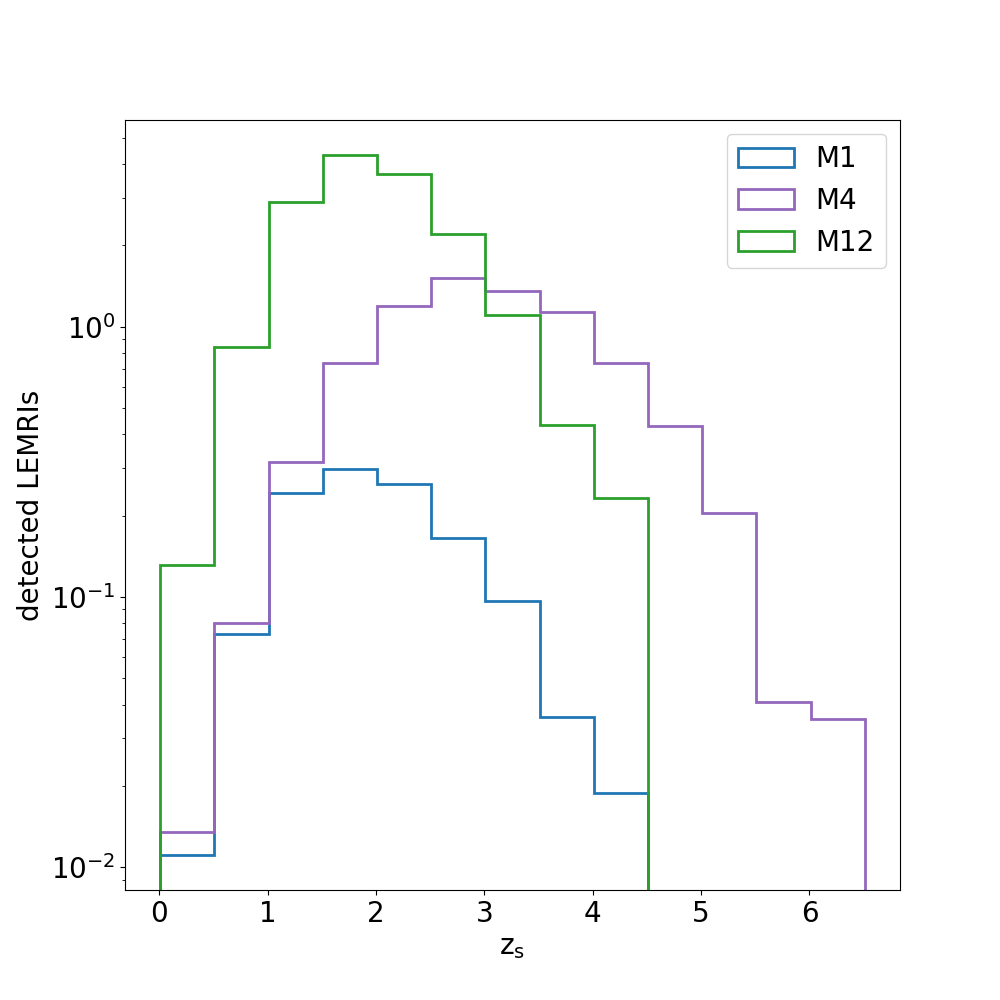}
    \caption{Observed LEMRIs in 4 years as a function of source redshift. We plot in blue M1, in violet M4 and in green M12, assuming no EMRI background confusion noise.}
    \label{fig:observd_lemri}
\end{figure}

Figure~\ref{fig:M1} shows the MBH mass/redshift distribution of the detected EMRIs for population model M1 at different SNR thresholds with or without the EMRI background noise.
Fig.~\ref{fig:M1M4M12} shows the MBH mass/redshift distribution of the detected EMRIs for three representative models M1, M4, and M12 in orange, violet and green respectively.\\
\indent Fig.~\ref{fig:dn_dz_20_nobg} shows the number of visible LEMRIs in the presence of magnification $\mu$ per redshift bin, $dz_{\rm S}$ for all 12 population models, computed for SNR threshold of 20, without considering the EMRI confusion noise.\\
\indent Fig.~\ref{fig:observd_lemri} shows the redshift distribution of the observed LEMRIs over a time of 4 years for M1, M4, M12, assuming a detection threshold of 20 and no EMRI confusion noise. For M1, our fiducial model, the number of LEMRIs above threshold peaks between $1\lesssim z \lesssim2$, which justifies our assumption of a LEMRI located at $z_{\rm S}=1.26$ in the parameter estimation section.


\begin{figure*}
    \includegraphics[width =\textwidth]{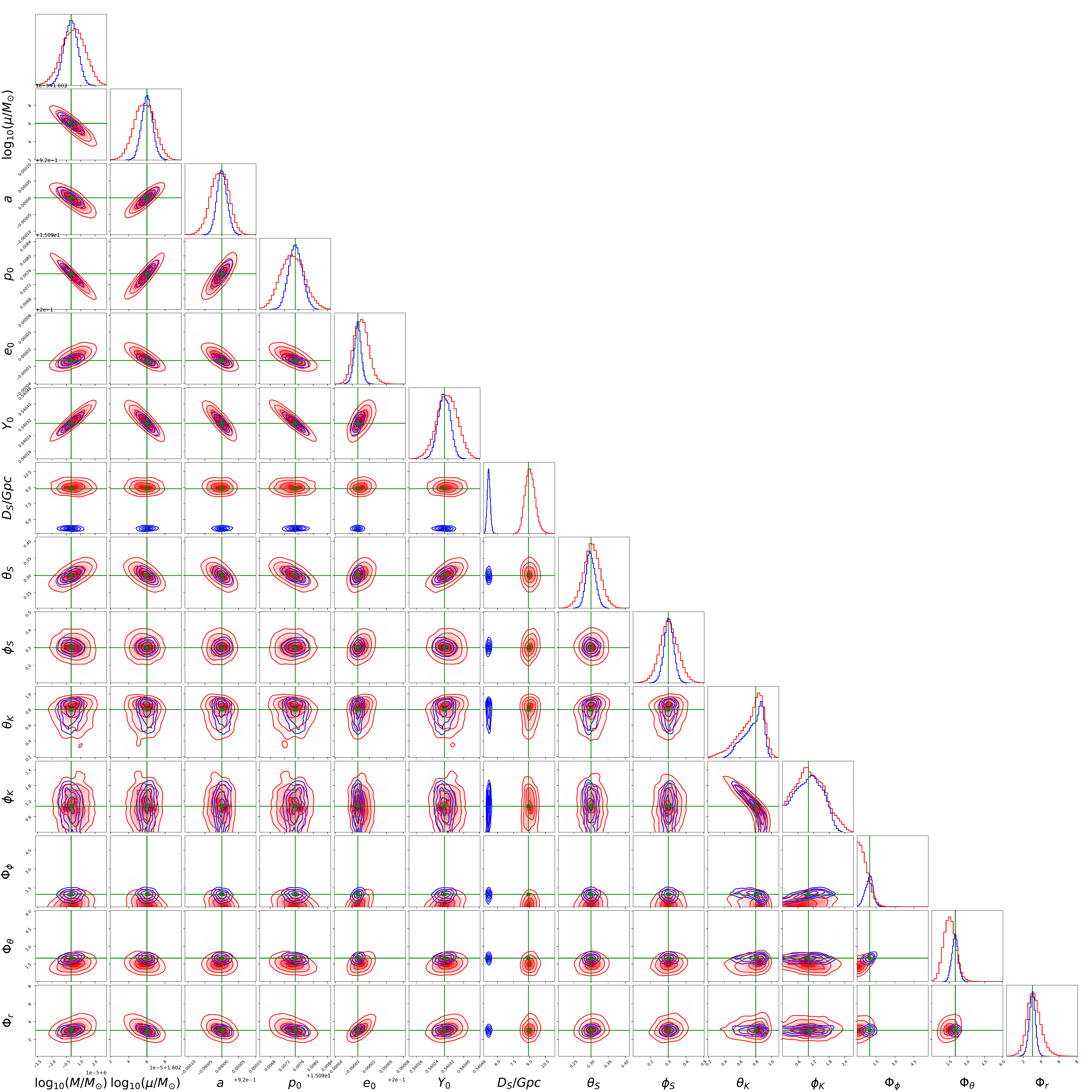}
    \caption{The blue posterior represents $p(\boldsymbol{\theta}^{E}|s,z_{S} = 1.26)$ with $\Delta t = 0$. The red posterior is the same except with a time shift $\Delta t = 45.6$ days. The black vertical line indicates the true values of the LEMRI. Due to our choice of impact factor $y = 0.5$, the effective distance $D^{-}$ is equal to the luminosity distance of the source $D_{S}$.}
    \label{fig:multiple_mismodel_PE}
\end{figure*}
\begin{figure*}
    \centering
    \includegraphics[width = \textwidth]{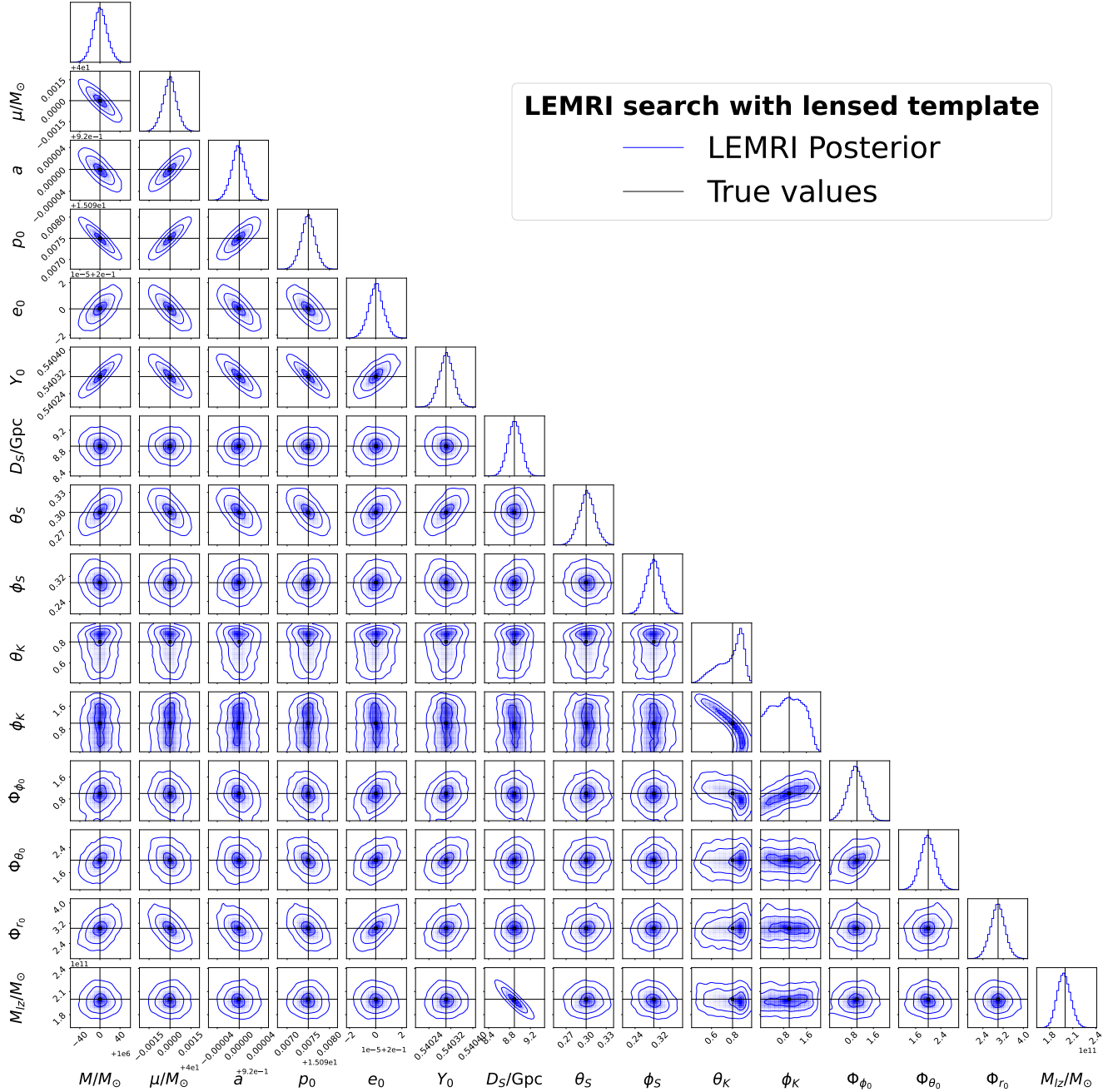}
    \caption{The blue posteriors are generated using lensed waveform models with SNR $\sim 44.98$. The green vertical line indicates the true parameters of the source. The last row is joint/marginal posteriors on the redshifted lensed mass $M_{Lz}$ recovered using the LEMRI.
    }
    \label{fig:LEMRI_PE_Plot}
\end{figure*}

\twocolumngrid

\nocite{*}
\bibliographystyle{apsrev}
\bibliography{main_V3}

\end{document}